\documentclass[aps, prd, twocolumn, nofootinbib,]{revtex4-1}

\usepackage[top=1in, bottom=1in, left=1in, right=1in]{geometry}

\usepackage{amsmath}
\usepackage{amssymb}
\usepackage{amsfonts}
\usepackage{wasysym}
\usepackage{graphicx}
\usepackage{comment}
\usepackage{tensor}

\def\filetype{pdf}
\def\path{}

\begin{document}


\title{Perturbing the ground state of Dirac stars}
\author{Emanuel Daka, Nhon N.\ Phan, and Ben Kain}
\affiliation{Department of Physics, College of the Holy Cross, Worcester, Massachusetts 01610, USA}

\begin{abstract}
\noindent Dirac stars are self-gravitating configurations of spin-1/2 fermions in which the fermions are described by the Dirac equation.  After a detailed review of the derivation of the equations and their static solutions, we present an in-depth dynamical stability analysis of the ground state similar to previous studies for boson stars.  We confirm that there exist both stable and unstable branches of static solutions and show that weakly perturbed Dirac stars from the unstable branch migrate to the stable branch.  We also show that strongly perturbed Dirac stars from the stable branch migrate to the stable branch if their mass is below a critical value.  If their mass is above the critical value they can migrate to the stable branch or collapse and form a black hole.  For strongly perturbed Dirac stars from the unstable branch we show that the addition of even a small amount of mass leads to collapse, while if we decrease their mass they migrate to the stable branch.
\end{abstract} 

\maketitle


\section{Introduction}

The Einstein-Dirac system is composed of the Dirac equation coupled to general relativity.  Spherically symmetric static solutions in this system were first found by Finster, Smoller, and Yau \cite{Finster:1998ws}.  Their solutions describe self-gravitating configurations of spin-1/2 fermions which are called Dirac stars.

The study of self-gravitating systems of matter analogous to Dirac stars is extensive.  The most heavily studied are boson stars, which are static solutions in the Einstein-Klein-Gordon system with a complex scalar field \cite{Schunck:2003kk, Liebling:2012fv}.  Related to boson stars are oscillatons \cite{Seidel:1991zh} in that they too are solutions in the Einstein-Klein-Gordon system but with a real instead of complex scalar field.  Oscillatons, however, are not static and the spacetime is time-dependent and oscillating.  Self-gravitating static solutions also exist with spin-1 fields:\  In the Einstein-Yang-Mills system, where the matter sector is made up of $SU(2)$ fields, the solutions are called Bartnik-McKinnon solutions \cite{Bartnik:1988am} and in the Einstein-Proca system, where the matter sector is made up of complex vector fields, the solutions are called Proca stars \cite{Brito:2015pxa}.    Upon combining different types of matter, there are even more possibilities, which include charged boson stars \cite{Jetzer:1989av}, gravitating magnetic monopoles \cite{VanNieuwenhuizen:1975tc, Ortiz:1991eu, Breitenlohner:1991aa}, etc.

Once solutions of this sort are found, an immediate question is are they stable?  In the case of boson stars, stability was studied in \cite{Gleiser:1988ih, Lee:1988av} where both stable and unstable branches for the solutions were identified.  This stability analysis was at the level of first order perturbations about the static solutions and therefore could not accommodate large perturbations or the evolution of growing instabilities.  A dynamical stability analysis that could accommodate these shortfalls was undertaken by Seidel, Suen, et al.\ \cite{Seidel:1990jh, Balakrishna:1997ej} where, among other things, they confirmed the stability of the stable solutions and showed that the unstable solutions migrate to stable solutions.  Their framework for a dynamical stability analysis was later used in a study of oscillatons \cite{Alcubierre:2003sx, UrenaLopez:2012zz} where similar results were found.  A similar framework was used for Proca stars in \cite{Sanchis-Gual:2017bhw}, which also found similar results.

In the original work on Dirac stars, Finster et al.\ made a semi-analytical stability analysis and determined that there exist both stable and unstable branches of solutions \cite{Finster:1998ws}.  Their stability analysis was  at the level of first order perturbations about the static solutions.  In this work we make a dynamical study of the stability of Dirac stars using the framework of Seidel and Suen \cite{Seidel:1990jh}.  Our results are similar to results found for boson stars \cite{Seidel:1990jh, Balakrishna:1997ej}, oscillatons \cite{Alcubierre:2003sx, UrenaLopez:2012zz}, and Proca stars \cite{Sanchis-Gual:2017bhw}.  Specifically, after identifying stable ($S$) and unstable ($U$) branches for Dirac stars, we dynamically evolve both weakly and strongly perturbed $S$- and $U$-branch solutions.  We corroborate that weakly perturbed $S$-branch solutions are stable and show that weakly perturbed $U$-branch solutions migrate to the $S$-branch.  If the mass of a strongly perturbed $S$-branch solution stays beneath the critical mass, which is the mass of the largest mass Dirac star, the system migrates to the $S$-branch.  If the mass is raised above the critical mass then the system can either migrate to the $S$-branch or collapse and form a black hole.  Strongly perturbed $U$-branch solutions migrate to the $S$-branch if the perturbation decreases their mass and collapses to form a black hole if their mass is raised as little as a few percent.  We focus exclusively on Dirac stars in the ground state and our work is entirely classical.  There exist also excited states for Dirac stars \cite{Finster:1998ws}.  A dynamical stability analysis for excited states will be presented elsewhere.

Our interest in Dirac stars is as a solitonic self-gravitating  classical solution to general relativity and with its relationship to comparative systems such as boson stars and Proca stars \cite{Herdeiro:2017fhv} (as a classical system, we are not making comparisons with, say, baryons in compact objects \cite{Glendenning:1997wn}).  Compared to boson stars, the study of Dirac stars is somewhat limited, though it is currently experiencing a resurgence of interest.  Dirac star solutions have been found in which the fermions are electrically charged \cite{Finster:1998ux, Bohun:2000ie} and gauged under $SU(2)$ \cite{Finster:2000ps}.  Dynamical solutions of the Einstein-Dirac system \cite{Ventrella:2003fu, Zeller:2006rm} were initiated by Ventrella and Choptuik \cite{Ventrella:2003fu} in a study of type II critical collapse.  More recently a comparison of the space of static solutions for boson, Dirac, and Proca stars was given in \cite{Herdeiro:2017fhv, Blazquez-Salcedo:2019qrz}, Dirac star solutions in which the fermions have interactions were found in \cite{Adanhounme, Dzhunushaliev:2018jhj, Dzhunushaliev:2019kiy}, and for the first time spinning Dirac stars were solved for in \cite{Herdeiro:2019mbz}.

In the next section we give a detailed review of the spherically symmetric Einstein-Dirac system.  This includes coupling spinors to curved space, the derivation of the Dirac spinor ansatz and energy-momentum tensor, and the scaling of fields to form dimensionless quantities.  In Sec.\ \ref{sec:static} we review spherically symmetric static solutions of the Einstein-Dirac system, i.e.\ Dirac stars, and in Sec.\ \ref{sec:dynamic} we present our dynamical stability analysis.  We conclude in Sec.\ \ref{sec:conclusion}.


\section{Einstein-Dirac}

In this section we derive the equations for the Einstein-Dirac system.  The equations will be  time-dependent and describe self-gravitating spin-1/2 fermions in spherical symmetry.  In the first subsection we review coupling spinors to curved space using the vierbein formalism.  Subsequent subsections derive the form of the fermion wave function we shall use, which is often called the Dirac spinor ansatz, and the energy-momentum tensor.  The final subsection discusses scaling and dimensionless variables used in our numerical studies.


\subsection{Spinors in curved space}

Coupling spinors to gravity is most commonly done in the vierbein formalism \cite{Weinberg:1972kfs, Carroll:2004st, Freedman:2012zz}, where the vierbein, $e\indices{_a^\mu}$, is defined by
\begin{equation}
g_{\mu \nu} = e\indices{_{a \mu}} e\indices{^a_\nu}, \qquad
\eta_{ab} = e\indices{_{a \mu}} e\indices{_b^\mu}.
\end{equation}
We use lowercase Greek letters for ``curved space indices" and lowercase Latin letters from the beginning of the alphabet for ``flat space indices."  $g_{\mu\nu}$ is the curved space metric of general relativity and $\eta_{ab}$ is the flat space (Minkowski) metric.  Curved space indices are raised and lowered with the curved space metric and flat space indices are raised and lowered with the flat space metric.  We adopt the mostly plus metric signature $(-,+,+,+)$ and use units such that $c=\hbar=1$, but retain the gravitational constant $G$.

The action for our system is
\begin{equation} \label{action}
S = \int d^4x \, e \left( \frac{R}{16\pi G} +  \mathcal{L} \right),
\end{equation}
where $R$ is the Ricci scalar, $e = \sqrt{-g}$ is the determinant of the vierbein, and $g$ is the determinant of the metric.  This action minimally couples the matter sector Lagrangian, $\mathcal{L}$, to gravity.  Our matter sector is composed of massive fermions with Lagrangian
\begin{equation} \label{L}
\mathcal{L} = \sum_x \left[ \frac{1}{2} \bar{\psi}_x \gamma^\mu \nabla_\mu \psi_x - \frac{1}{2} (\nabla_\mu \bar{\psi}_x) \gamma^\mu \psi_x 
- \mu \bar{\psi}_x \psi_x \right],
\end{equation}
where $\psi_x$ is a four-component Dirac spinor and $\bar{\psi}_x$ is its adjoint.  We have allowed for the possibility of more than one fermion, labeled by $x$, with common mass $\mu$.  We will see below that spherical symmetry requires at least two fermions.  

$\gamma^\mu$ are curved space $\gamma$-matrices and are related to flat space $\gamma$-matrices, $\gamma^a$, through the vierbein:
\begin{equation}
\gamma^\mu = e\indices{_a^\mu} \gamma^a, \qquad
\gamma^a = e\indices{^a_\mu} \gamma^\mu.
\end{equation}
$\gamma$-matrices are defined as usual by their anti-commutation relations:
\begin{equation}
\{\gamma^\mu, \gamma^\nu\} = 2 g^{\mu\nu}, \qquad
\{\gamma^a, \gamma^b\} = 2\eta^{ab}.
\end{equation}
The specific representations we choose for the $\gamma^a$-matrices, vierbein, metric, and adjoint spinor are given in the next subsection.

The covariant derivatives in (\ref{L}) are defined by
\begin{equation}
\begin{split}
\nabla_\mu \psi_x &= \partial_\mu \psi_x + \frac{1}{4} \omega\indices{_\mu^{ab}} \gamma_{ab}  \psi_x
\\
\nabla_\mu \bar{\psi}_x &= \partial_\mu \bar{\psi}_x - \frac{1}{4} \bar{\psi}_x \omega\indices{_\mu^{ab}} \gamma_{ab},
\end{split}
\end{equation}
where $\gamma_{ab} \equiv \gamma_{[a}\gamma_{b]} = [\gamma_a,\gamma_b]/2$ and $\omega\indices{_\mu^{ab}}$ is the spin connection:
\begin{align}
w_{\mu ab} &= 
\frac{1}{2} e\indices{_a^\alpha} (\partial_\mu e_{b\alpha} - \partial_\alpha e_{b\mu} )
+ \frac{1}{2} e\indices{_b^\beta}
(\partial_\beta e_{a\mu} - \partial_\mu e_{a\beta} )
\notag
\\
&\qquad - \frac{1}{2} 
e\indices{^c_\mu}
e\indices{_a^\alpha}
e\indices{_b^\beta}  (\partial_\alpha e_{c\beta} - \partial_\beta e_{c\alpha} ).
\end{align}
The spin connection plays a role similar to the Christoffel connection, but for spinors and objects with flat space indices.  It can be written in terms of the Christoffel connection,
\begin{equation}
\omega_{\mu a b} = e_{a \nu}e\indices{_b^\lambda} \Gamma^\nu_{\mu\lambda} - e\indices{_b^\lambda}\partial_\mu e_{a \lambda},
\end{equation}
which may be a more efficient way of computing it if the Christoffel connection is known.

The Dirac equation is obtained as the equations of motion of the Lagrangian in (\ref{L}):
\begin{equation} \label{Dirac eq}
(\gamma^\mu \nabla_\mu - \mu)\psi_x = 0, \qquad
(\nabla_\mu \bar{\psi}_x)\gamma^\mu + \mu \bar{\psi}_x = 0.
\end{equation}
Since the Lagrangian possess a global $U(1)$ symmetry for each fermion, $\psi_x \rightarrow e^{i\beta_x} \psi_x$, where $\beta_x$ is a constant, there exists a conserved current for each fermion,
\begin{equation} \label{current}
j^\mu_x = -i \bar{\psi}_x \gamma^\mu \psi_x,
\end{equation}
with associated conserved charge
\begin{equation} \label{Q def}
Q_x = \int_\Sigma d^3x \, \sqrt{\gamma} (-n_\mu j^\mu_x),
\end{equation}
where $n_\mu$ is a timelike unit vector normal to the spatial slice $\Sigma$ and $\gamma$ is the determinant of the spatial metric on $\Sigma$.  The conserved charge gives the particle number for each type of fermion.

The Einstein field equations, 
\begin{equation}
G_{\mu\nu} = 8\pi G T_{\mu\nu},
\end{equation}
where $G_{\mu\nu}$ is the Einstein tensor and $T_{\mu\nu}$ is the energy-momentum tensor, are obtained by variation of the action (\ref{action}) with respect to the vierbein or equivalently variation with respect to the metric.  This variation gives for the energy-momentum tensor
\begin{equation} \label{energy-momentum sum}
T_{\mu\nu} = \sum_x T_{\mu\nu}^x,
\end{equation}
where
\begin{equation} \label{energy-momentum}
\begin{split}
T_{\mu\nu}^x &= 
- \frac{1}{4}
\Bigl[
\bar{\psi}_x \gamma_\mu \nabla_\nu \psi_x
+ \bar{\psi}_x \gamma_\nu \nabla_\mu \psi_x
\\
&\qquad
- (\nabla_\mu \bar{\psi}_x)\gamma_\nu  \psi_x
- (\nabla_\nu \bar{\psi}_x)\gamma_\mu \psi_x
\Bigr].
\end{split}
\end{equation}
For a pedagogical derivation of the Dirac energy-momentum tensor see, for example, \cite{Shapiro:2016pfm, Freedman:2012zz}.


\subsection{Representations}

The equations in the previous subsection were written down without choosing a specific form, or representation, for quantities such as the vierbein and $\gamma$-matrices.  Making a choice of representation helps in deriving the system of equations that we will solve numerically.  In the following, when specifying explicit components, we use $a=0,1,2,3$ and $\mu=t,r,\theta,\phi$.  

We adopt the conventions used by Ventrella and Choptuik \cite{Ventrella:2003fu} for the metric, the flat and curved space $\gamma$-matrices, the vierbein, and the adjoint spinor.  Specifically, we take the metric to have the spherically symmetric form
\begin{equation} \label{metric}
\begin{split}
ds^2 &= -\alpha^2(t,r) dt^2 + a^2(t,r) dr^2 
\\
&\qquad + r^2 d\theta^2 + r^2  \sin^2\theta d\phi^2,
\end{split}
\end{equation}
where $\alpha(t,r)$ and $a(t,r)$ are metric functions that we solve for using the Einstein field equations.  This is not the most general spherically symmetric form for the metric \cite{AlcubierreBook, BaumgarteBook}, but it is a particularly simple form and convenient for our purposes.  

For flat space $\gamma^a$-matrices we use the Dirac representation:
\begin{equation}
\gamma^0 = i
\begin{pmatrix}
1 & 0 \\ 0 & -1
\end{pmatrix},
\qquad
\gamma^j = i
\begin{pmatrix}
0 & \sigma^j \\ -\sigma^j & 0
\end{pmatrix},
\end{equation}
where $j=1,2,3$ and where the $\sigma^j$ are the standard Pauli matrices:
\begin{equation}
\sigma^1 = 
\begin{pmatrix}
0 & 1 \\ 1 & 0
\end{pmatrix},
\quad
\sigma^2 = 
\begin{pmatrix}
0 & -i \\ i & 0
\end{pmatrix}, 
\quad
\sigma^3 = 
\begin{pmatrix}
1 & 0 \\ 0 & -1
\end{pmatrix}.
\end{equation}
We take the curved space $\gamma^\mu$-matrices to be given by
\begin{equation} \label{gamma mu a}
\gamma^t = \frac{\gamma^0}{\alpha}, \quad
\gamma^r = \frac{\gamma^3}{a}, \quad
\gamma^\theta = \frac{\gamma^2}{r}, \quad
\gamma^\phi = \frac{\gamma^1}{r\sin\theta},
\end{equation}
which fixes the vierbein through $\gamma^\mu = e\indices{_a^\mu} \gamma^a$.  The particular choice in (\ref{gamma mu a}) of associating the angular components $\theta$ and $\phi$ with the off-diagonal Pauli matrices $\sigma^1$ and $\sigma^2$ is not strictly necessary (we could have, for example, taken the vierbein to be diagonal), but it simplifies separating out the angular dependence in the Dirac equation, which we do in the next subsection.  

Lastly, we take the adjoint spinor to be defined by
\begin{equation}
\bar{\psi} = \psi^\dag \beta, \qquad \beta = -i\gamma^0,
\end{equation}
where $\beta$ is called the Hermitizing matrix.  We note that the overall sign of the Lagrangian in (\ref{L}) and the conserved current in (\ref{current})  assumed this definition.


\subsection{Equations of motion and the ansatz}

In this subsection we find a form for the Dirac spinor that is consistent with the Dirac equation in (\ref{Dirac eq}) and the spherically symmetric metric in (\ref{metric}).  The standard approach is to follow Unruh \cite{Unruh:1973bda} and Chandrasekhar \cite{Chandrasekhar:1976ap, Chandrasekhar:1985kt} and look for separable solutions:
\begin{equation}
\psi = 
\begin{pmatrix}
\psi_1 (t,r,\theta,\phi) \\
\psi_2 (t,r,\theta,\phi) \\
\psi_3 (t,r,\theta,\phi) \\
\psi_4 (t,r,\theta,\phi)
\end{pmatrix}
=
\begin{pmatrix}
R_1(t,r) \Theta_1(\theta,\phi) \\
R_2(t,r) \Theta_2(\theta,\phi) \\
R_3(t,r) \Theta_3(\theta,\phi) \\
R_4(t,r) \Theta_4(\theta,\phi)
\end{pmatrix}.
\end{equation}
Note that the $R$'s and $\Theta$'s are in general complex.  Plugging this into the Dirac equation in (\ref{Dirac eq}) and using the metric in (\ref{metric}), we end up with the following four equations:
\begin{widetext}
\begin{equation} \label{four eom}
\begin{split}
\frac{ir}{\alpha} \left(\frac{\dot{R}_1}{R_1} + \frac{\dot{a}}{2a} \right)
\frac{R_1}{R_4} \frac{\Theta_1}{\Theta_3}
+ \frac{ir}{a} \left(\frac{R_3'}{R_3} + \frac{\alpha'}{2\alpha} + \frac{1}{r} \right) 
\frac{R_3}{R_4} 
-\mu r \frac{R_1}{R_4} \frac{\Theta_1}{\Theta_3}
 &=  - \left( \frac{\partial_\theta \Theta_4}{\Theta_4} + \frac{\cot\theta}{2} \right) 
\frac{\Theta_4}{\Theta_3}
- \frac{i}{\sin\theta} \frac{\partial_\phi \Theta_4} {\Theta_4} 
\frac{\Theta_4}{\Theta_3}
\\
\frac{ir}{\alpha} \left(\frac{\dot{R}_2}{R_2} + \frac{\dot{a}}{2a} \right)
\frac{R_2}{R_3} \frac{\Theta_2}{\Theta_4}
- \frac{ir}{a} \left(\frac{R_4'}{R_4} + \frac{\alpha'}{2\alpha} + \frac{1}{r} \right) \frac{R_4}{R_3} 
-\mu r \frac{R_2}{R_3} \frac{\Theta_2}{\Theta_4}
&=  \left( \frac{\partial_\theta \Theta_3}{\Theta_3} + \frac{\cot\theta}{2} \right) 
\frac{\Theta_3}{\Theta_4}
- \frac{i}{\sin\theta} \frac{\partial_\phi \Theta_3} {\Theta_3} \frac{\Theta_3}{\Theta_4}
\\
\frac{ir}{\alpha} \left(\frac{\dot{R}_3}{R_3} + \frac{\dot{a}}{2a} \right)
\frac{R_3}{R_2} \frac{\Theta_3}{\Theta_1}
+ \frac{ir}{a} \left(\frac{R_1'}{R_1} + \frac{\alpha'}{2\alpha} + \frac{1}{r} \right) \frac{R_1}{R_2} 
+\mu r \frac{R_3}{R_2} \frac{\Theta_3}{\Theta_1}
&= 
-\left( \frac{\partial_\theta \Theta_2}{\Theta_2} + \frac{\cot\theta}{2} \right)
 \frac{\Theta_2}{\Theta_1}
- \frac{i}{\sin\theta} \frac{\partial_\phi \Theta_2} {\Theta_2}
\frac{\Theta_2}{\Theta_1}
\\
 \frac{ir}{\alpha} \left(\frac{\dot{R}_4}{R_4} + \frac{\dot{a}}{2a} \right)
 \frac{R_4}{R_1} \frac{\Theta_4}{\Theta_2}
- \frac{ir}{a} \left(\frac{R_2'}{R_2} + \frac{\alpha'}{2\alpha} + \frac{1}{r} \right)
\frac{R_2}{R_1} 
+\mu r \frac{R_4}{R_1} \frac{\Theta_4}{\Theta_2}
&=
  \left( \frac{\partial_\theta \Theta_1}{\Theta_1} + \frac{\cot\theta}{2} \right) 
\frac{\Theta_1}{\Theta_2}
- \frac{i}{\sin\theta} \frac{\partial_\phi \Theta_1} {\Theta_1} \frac{\Theta_1}{\Theta_2},
\end{split}
\end{equation}
\end{widetext}
where a dot denotes a $t$-derivative and a prime denotes an $r$-derivative.  To complete the separation of variables we assume
\begin{equation} \label{R theta assumptions}
R_2 = i R_1, \quad
R_4 = i R_3, \quad
\Theta_3 = \Theta_1, \quad
\Theta_4 = -\Theta_2,
\end{equation}
which reduces the above four equations to two independent equations.

We focus on the resulting angular equations first.  They can be written as
\begin{equation}
\eth_+^{(-1/2)} \Theta_2 = - n \Theta_1, \qquad
\eth_-^{(+1/2)} \Theta_1 = n \Theta_2,
\end{equation}
where $n$ is the separation constant and
\begin{equation}
\eth_\pm^{(s)}
= \mp \frac{i}{\sin\theta} \partial_\phi
 - \partial_\theta
 \pm s \cot\theta.
\end{equation}
$\eth_+^{(s)}$ is the raising operator and $\eth_-^{(s)}$ is the lowering operator for spin-weighted spherical harmonics, ${_s Y_{\ell m}}$, of spin weight $s$:
\begin{equation}
\eth_\pm^{(s)}  ({_s Y_{\ell m}}) = \pm \sqrt{(\ell \mp s)(\ell \pm s + 1)} ( {_{s\pm 1} Y_{\ell m}}).
\end{equation} 
We thus find that $\Theta_1 = {_{+1/2}Y_{\ell m}}$ and $\Theta_2 = {_{-1/2}Y_{\ell m}}$ are spin-weighted spherical harmonics.  We will find below that having only one spin-1/2 fermion violates spherical symmetry and that to preserve spherical symmetry we need two (or more) fermions.  We consider only two fermions and  use $\Theta_1 = {_{1/2}Y_{(1/2)(1/2)}}$ and $\Theta_2 = {_{-1/2}Y_{(1/2)(1/2)}}$ for one fermion and $\Theta_1 = {_{1/2}Y_{(1/2)(-1/2)}}$ and $\Theta_2 = {_{-1/2}Y_{(1/2)(-1/2)}}$ for the other fermion \cite{Finster:1998ws, Ventrella:2003fu}, where the spin-weighted spherical harmonics are given by
\begin{equation} \label{spin-weighted}
\begin{split}
{_{\pm {1/2}}Y_{(1/2)(1/2)}} &= \frac{1}{2\sqrt{\pi}} e^{i\phi/2} y_\pm (\theta),
\\
{_{\pm {1/2}}Y_{(1/2)(-1/2)}} &= \pm \frac{1}{2\sqrt{\pi}} e^{-i\phi/2} y_\mp(\theta),
\end{split}
\end{equation}
with
\begin{equation}
y_\pm(\theta) \equiv \sqrt{\frac{1 \mp \cos\theta}{2}} = 
\begin{array}{c}
\sin(\theta/2) \\ \cos(\theta/2)
\end{array}.
\end{equation}
Our choice of spin-weighted spherical harmonics fixes the separation constant to
\begin{equation} \label{separation constant}
n = -1.
\end{equation}

Recall that the four equations in (\ref{four eom}) were reduced to two independent equations by assuming (\ref{R theta assumptions}).  Having fixed the separation constant in (\ref{separation constant}) we can now write down the radial equations of motion.  Before doing so it is convenient to define the complex functions $F(t,r)$ and $G(t,r)$ as
\begin{equation}
R_1(t,r) \equiv \frac{F(t,r)}{r\sqrt{a(t,r)}}, 
\qquad
R_2(t,r) \equiv \frac{G(t,r)}{r\sqrt{a(t,r)}}.
\end{equation}
Extracting out the factors of $1/r\sqrt{a}$ removes an inconvenient time derivative that would otherwise be in the radial equations of motion, which are now given by
\begin{equation} \label{radial eom}
\begin{split}
\dot{F} &= - \frac{\alpha G}{r}
\left[ 1 + \frac{r}{a} 
\left(\frac{G'}{G} - \frac{a'}{2a} + \frac{\alpha'}{2\alpha} \right) 
+ i\mu r \frac{F}{G} \right],
\\
\dot{G} &= \frac{\alpha F}{r}
\left[ 1 - \frac{r}{a} 
\left(\frac{F'}{F} - \frac{a'}{2a} + \frac{\alpha'}{2\alpha}  \right)
+ i\mu {r} \frac{G}{F}
\right].
\end{split}
\end{equation}

We end this subsection by giving the final form of our Dirac spinors:
\begin{equation} \label{psi pm}
\psi_\pm = 
\frac{e^{\pm i \phi/2}}{2 r \sqrt{\pi a(t,r)}}
\begin{pmatrix}
F(t,r) y_\pm(\theta) \\
\pm iF(t,r) y_\mp(\theta) \\
G(t,r) y_\pm(\theta) \\
\mp iG(t,r) y_\mp(\theta)
\end{pmatrix},
\end{equation}
where the index $x$, which labeled the particular fermion, has been replaced with $\pm$.  The equation above is often referred to as the Dirac spinor ansatz for spherical symmetry.  Though it does not look the same as forms found in the literature \cite{Finster:1998ws, Ventrella:2003fu, Herdeiro:2017fhv}, we show in Appendix \ref{app:ansatz} that it is equivalent. 


\subsection{Energy-momentum tensor}

The equation for the energy-momentum tensor of an individual fermion, $T_{\mu\nu}^x$, is given in (\ref{energy-momentum}).  For a time-dependent energy-momentum tensor to be spherically symmetric its only nonvanishing components can be the diagonal components and $T_{tr} = T_{rt}$.  The components of the energy-momentum tensor for the Dirac spinor $\psi_\pm$ in (\ref{psi pm}) has nonvanishing $T_{t\phi}^\pm$ and $T_{r\phi}^\pm$ and thus breaks spherical symmetry.  The total energy-momentum tensor,
\begin{equation}
T_{\mu\nu} = T_{\mu\nu}^+ + T_{\mu\nu}^-,
\end{equation}
however, is spherically symmetric.  In this way we can understand why two (or more) fermions are necessary to preserve spherical symmetry.  The nonvanishing components of the spherically symmetric energy-momentum tensor are
\begin{equation} \label{energy-momentum comps}
\begin{split}
T_{tt} &=
 \frac{-\alpha}{2\pi r^2 a} \text{Im} ( F^*\dot{F} + G^*\dot{G})
\\
T_{tr}
&= 
 \frac{-1}{4\pi r^2}
\text{Im} \left[ 
\frac{\alpha}{a} ( F^*F' + G^*G')
- (\dot{F}G^* + F^* \dot{G}) \right]
\\
T_{rr}
&= \frac{1}{ 2\pi r^2} \text{Im}
(F'G^* + F^* G')
\\
T_{\theta\theta}
&=  \frac{1}{ 2\pi r a} \text{Im}
 ( F^* G) 
\\
T_{\phi\phi}
&= T_{\theta\theta} \sin^2\theta.
\end{split}
\end{equation}


\subsection{Equations}

We can now list the complete set of equations.  In doing this we decompose the complex fermion functions $F$ and $G$ into their real and imaginary parts:
\begin{equation}
\begin{split}
F(t,r) &= F_1(t,r) + i F_2(t,r), 
\\
G(t,r) &= G_1(t,r) + i G_2(t,r).
\end{split}
\end{equation}
The evolution equations for the matter sector come from the radial equations of motion in (\ref{radial eom}), which can be written
\begin{equation} \label{matter eom}
\begin{split}
\dot{F}_1 
&= -\frac{\alpha}{r} \left( G_1 - \mu r F_2 \right)
- \sqrt{\frac{\alpha}{a}}
\partial_r \left( \sqrt{\frac{\alpha}{a}} G_1 \right)
\\
\dot{F}_2
&= -\frac{\alpha}{r} \left(G_2 + \mu r F_1 \right)
- \sqrt{\frac{\alpha}{a}}
\partial_r \left( \sqrt{\frac{\alpha}{a}} G_2 \right)
\\
\dot{G}_1 
&= \frac{\alpha}{r} \left( F_1 - \mu r G_2 \right)
- \sqrt{\frac{\alpha}{a}}
\partial_r \left( \sqrt{\frac{\alpha}{a}} F_1 \right)
\\
\dot{G}_2 
&= \frac{\alpha}{r} \left( F_2 + \mu r G_1 \right)
- \sqrt{\frac{\alpha}{a}}
\partial_r \left( \sqrt{\frac{\alpha}{a}} F_2 \right).
\end{split}
\end{equation}
The metric functions $a(t,r)$ and $\alpha(t,r)$ obey the constraint equations \cite{AlcubierreBook, BaumgarteBook}
\begin{equation} \label{alpha a eqs}
\begin{split}
\frac{\alpha'}{\alpha} &= 4\pi G r  a^2 S\indices{^r_r} + \frac{a^2-1}{2r}  \\
\frac{a'}{a} &= 
4\pi G r a^2\rho
- \frac{a^2 - 1}{2r},
\end{split}
\end{equation}
which follow from the Einstein field equations, where the energy density, $\rho(t,r)$, and the stress, $S\indices{^r_r}(t,r)$, follow from the energy-momentum tensor and are given by
\begin{align}
\rho &= 
\frac{1}{2\pi r^2 a^2} 
\biggl[  a \mu  (F_1^2 + F_2^2 - G_1^2 - G_2^2)
\notag \\
&\qquad\qquad
+\frac{2 a}{r} ( F_1 G_2 - F_2 G_1 )
\label{rho Srr}
\\
&\qquad\qquad
+  F_1 G_2'-  F_2 G_1'+ G_1 F_2'- G_2 F_1'
\biggr]
\notag \\
S\indices{^r_r} &= \frac{1}{2\pi r^2 a^2 } ( F_1 G_2'-  F_2 G_1'+ G_1 F_2'- G_2 F_1' ).
\notag
\end{align}
In the following sections we will solve these equations numerically for both static and dynamic solutions.  Our dynamic solutions will be used to study the stability of the static solutions.

The metric in (\ref{metric}) is of the Schwarzschild form, motivating us to write $a^{-2}(t,r) = 1 - 2Gm(t,r)/r$.
Writing the bottom equation in (\ref{alpha a eqs}) in terms of $m(t,r)$ gives
\begin{equation} \label{m eq}
m' = 4\pi r^2 \rho.
\end{equation}
This tells us that we can interpret
\begin{equation} \label{mass function}
m(t,r) = \frac{r}{2 G}\left[1 - \frac{1}{a^2(t,r)}\right]
\end{equation}
as the total mass inside a radius $r$ and that the large $r$ limit gives the total integrated energy, i.e.\ the ADM mass:
\begin{equation}
M = \lim_{r\rightarrow \infty} m(t,r).
\end{equation}
We shall make frequent use of these mass equations.

The equation for the conserved charge, $Q_\pm$, in (\ref{Q def}) depends on the normal vector $n^\mu$ and the spatial metric $\gamma$.  For our metric (\ref{metric}), these are $n^\mu = (-\alpha,0,0,0)$ and $\gamma = a^2 r^4 \sin^2\theta$ \cite{AlcubierreBook, BaumgarteBook}.   Subbing in the Dirac spinor in (\ref{psi pm}) we find the same answer for $\psi_+$ and $\psi_-$, and thus drop the $\pm$ on $Q$:
\begin{equation} \label{Q def 2}
\begin{split}
Q &= \int \left(|F|^2 + |G|^2 \right) dr
\\
&=  \int \left( F_1^2 + F_2^2 + G_1^2 + G_2^2 \right) dr,
\end{split}
\end{equation}
where we recall that $Q$ is equal to the particle number for each type of fermion.


\subsection{Scaling}

In numerical work it is important to use dimensionless quantities:
\begin{equation} \label{first scaling}
\begin{gathered}
\bar{r} \equiv \mu \, r, \qquad
\bar{t} \equiv \mu \, t,
\\
\overline{F}_{1,2} \equiv \sqrt{G \mu}\, F_{1,2}, \qquad
\overline{G}_{1,2} \equiv \sqrt{G \mu}\, G_{1,2},
\\
\bar{\rho} \equiv (4\pi G/\mu^2) \rho, \qquad
\overline{S}\indices{^r_r} \equiv (4\pi G/\mu^2) S\indices{^r_r},
\end{gathered}
\end{equation}
where bars indicate the dimensionless versions and the factors of $4\pi$ are included for convenience.  With this scaling, the fermion mass, $\mu$, is absorbed into the coordinates and fields and does not have to be specified.  The total mass inside a radius $r$, $m(t,r)$, scales as
\begin{equation}
\bar{m} \equiv (\mu/ m_\text{Pl}^2) m,
\end{equation}
where $m_\text{Pl} = 1/\sqrt{G}$ is the Planck mass.  The ADM mass, $M$, which is given by $m(t,r)$ in the limit $r\rightarrow \infty$, scales identically:
\begin{equation} \label{ADM scaling}
\overline{M} =  (\mu/ m_\text{Pl}^2) M.
\end{equation}

An alternative scaling \cite{Finster:1998ws, Herdeiro:2017fhv}, which we will occasionally present, is obtained as follows.  The equation for the conserved charge, $Q$, is given in (\ref{Q def 2}).  Using the scaling in (\ref{first scaling}) this becomes
\begin{equation}
\overline{Q} =  \int (|\overline{F}|^2 + |\overline{G}|^2) d\bar{r},
\end{equation}
where
\begin{equation} \label{Qbar}
\overline{Q} \equiv  (\mu/ m_\text{Pl})^2 Q.
\end{equation}
From (\ref{Qbar}) and (\ref{ADM scaling}) we have
\begin{equation} \label{Q ADM}
\frac{\mu}{m_\text{Pl}} = (\overline{Q}/Q)^{1/2}, \qquad
\frac{M}{m_\text{Pl}} = (Q/\overline{Q})^{1/2} \overline{M}.
\end{equation}
The (unscaled) physical charge $Q$ represents the number of each type of fermion.  Classically $Q$ can be any positive number.  In \cite{Herdeiro:2017fhv} it was argued that quantization enforces $Q=1$.  Although our treatment is entirely classical we will at times review results for $Q=1$.  When doing so, we will use tildes to indicate dimensionless quantities.  Upon setting $Q=1$ (\ref{Q ADM}) becomes
\begin{equation}
\begin{split}
\tilde{\mu} &\equiv \frac{\mu}{m_\text{Pl}} \Biggr|_{Q=1} = \overline{Q}^{1/2},
\\
\widetilde{M} &\equiv \frac{M}{m_\text{Pl}} \Biggr|_{Q=1} = \frac{\overline{M}}{\overline{Q}^{1/2}}.
\end{split}
\end{equation}


\section{Static solutions}
\label{sec:static}

In this section we review static solutions.   Static solutions were first found in \cite{Finster:1998ws} and have been studied by various authors \cite{Finster:1998ux, Bohun:2000ie, Finster:2000ps, Herdeiro:2017fhv, Adanhounme, Dzhunushaliev:2018jhj, Dzhunushaliev:2019kiy, Blazquez-Salcedo:2019qrz}.  By static we mean only that the spacetime is time-independent (the matter fields may retain a time-dependence).  For the spacetime to be time-independent, the components of the energy-momentum tensor in (\ref{energy-momentum comps}) must be time-independent and $T_{tr}$ must vanish.  This can be accomplished with matter fields of the form
\begin{equation}
F(t,r) = f(r) e^{-i\omega t}, \qquad
G(t,r) = ig(r) e^{-i\omega t},
\end{equation}
where $f(r)$ and $g(r)$ are real functions and $\omega$ is a real constant.  We assume $\omega$ is positive since, as shown below, this leads to a positive energy density.  The matter sector of static solutions is then composed of the two real functions $f(r)$ and $g(r)$ and the solutions can be labeled by $\omega$.

When constructing static solutions it is convenient to trade the metric functions $a(r)$ and $\alpha(r)$, which are now time-independent, for the mass function $m(r)$ in (\ref{mass function}) and
\begin{equation}
\sigma(r) \equiv a(r) \alpha(r).
\end{equation}
Constraint equations for $\alpha$, $a$, and $m$ are given in (\ref{alpha a eqs}) and (\ref{m eq}) and  evolution equations for $F$ and $G$ are given in (\ref{radial eom}).  Moving to the time-independent functions $m(r)$, $\sigma(r)$, $f(r)$, and $g(r)$, we have
\begin{equation} \label{static eqs}
\begin{split}
m' &= 4\pi r^2 \rho
\\
\frac{\sigma'}{\sigma} &= \frac{4\pi G r}{N} \left( \rho + S\indices{^r_r} \right )
\\
f' &=  f \left(\frac{Z}{2} + \frac{1}{r\sqrt{N}} \right)
- g\left(\frac{\mu}{\sqrt{N}}  + \frac{\omega}{N \sigma} \right)  
\\
g' &=  g \left(\frac{Z}{2} - \frac{1}{r\sqrt{N}}  \right)
- f \left(\frac{\mu}{\sqrt{N}} - \frac{\omega}{N \sigma} \right) ,
\end{split}
\end{equation}
where
\begin{equation} \label{static equations}
\begin{split}
N &= 1 - \frac{2 G m}{r}
\\
\rho 
&= \frac{\omega}{2\pi r^2 \sigma}\left( f^2 + g^2 \right)
\\
S\indices{^r_r} 
&= \frac{N}{2\pi r^2}\left( fg' - f' g \right)
\\
Z 
&= \frac{4\pi G r}{N} \left( \rho - S\indices{^r_r} \right) - \frac{2 G m}{r^2 N}.
\end{split}
\end{equation}
Note that $S\indices{^r_r}$ depends on $fg' - f'g$, which is easily written in terms of un-differentiated fields using the formulas for $f'$ and $g'$ in (\ref{static eqs}):
\begin{equation}
fg' - f' g 
=
\frac{\omega(f^2 + g^2)}{N\sigma}
- \frac{2f g}{r\sqrt{N}} - \frac{\mu(f^2 - g^2)}{\sqrt{N}}.
\end{equation}
As promised, the energy density, $\rho$ in (\ref{static equations}), is positive for positive $\omega$.

Once appropriate boundary conditions are identified, the equations above can be solved numerically using standard integration techniques.  Being a numerical solution, we move to the dimensionless variables defined in (\ref{first scaling}) along with the additional dimensionless quantity
\begin{equation}
\bar{\omega} \equiv \omega/\mu.
\end{equation}
Inner boundary conditions near $\bar{r}=0$ are found by expanding the fields in a power series, plugging the power series  into (\ref{static eqs}), and then equating coefficients of like powers.  We find the solutions
\begin{gather} 
\bar{m} = O(\bar{r}^3), \quad
\sigma = \sigma_0 +  \frac{1}{3} \bar{f}_1^2 (4\bar{\omega} - \sigma_0) \bar{r}^2 + O(\bar{r}^4),
\notag
\\
\bar{f} = \bar{f}_1 \bar{r} + O(\bar{r}^3), \quad
\bar{g}_2 =  \frac{\bar{f}_1(\bar{\omega} - \sigma_0)}{3\sigma_0} \bar{r}^2 + O(\bar{r}^4),
\label{inner BC}
\end{gather} 
which are parameterized in terms of the two unknown constants $\sigma_0$ and $\bar{f}_1$.  Outer boundary conditions for $\bar{r}\rightarrow\infty$ can be determined in two ways.  First, we require $\bar{r}^2 \bar{\rho} \rightarrow 0$ as $r\rightarrow \infty$ so that the total integrated energy is finite, which requires $\bar{f},\bar{g}\rightarrow 0$.  Second, we assume the spacetime is asymptotically Schwarzschild and thus $\sigma \rightarrow 1$.

To integrate the system of equations in (\ref{static eqs}) outward from some small $\bar{r}_\text{min}$, three constants must be specified:\ $\sigma_0$, $\bar{f}_1$, and $\bar{\omega}$.  A look at the various equations shows, however, that this can be reduced to two constants by defining
\begin{equation}
\hat{\sigma}(r) \equiv \sigma(r)/\sigma_0, \qquad \hat{\omega} \equiv \bar{\omega}/\sigma_0,
\end{equation}
and then replacing all $\sigma$ and $\bar{\omega}$ in favor of $\hat{\sigma}$ and $\hat{\omega}$.  The inner and outer boundary conditions for $\hat{\sigma}$ are, respectively,
\begin{equation} \label{inner BC 2}
\hat{\sigma} = 1 + \frac{1}{3} \bar{f}_1^2 (4\hat{\omega} - 1) \bar{r}^2 + O(\bar{r}^4), \qquad
\hat{\sigma} \rightarrow 1/\sigma_0.
\end{equation}
Previously we knew the outer boundary condition of $\sigma$ and parameterized its inner boundary condition in terms of the unknown constant $\sigma_0$.  By moving to $\hat{\sigma}$ we have switched this, so that the unknown constant $\sigma_0$ now parameterizes the outer boundary condition.  The value in doing this is that now only two constants, $\bar{f}_1$ and $\hat{\omega}$, must be specified at the inner boundary.

We can now solve for static solutions using the shooting method.  We begin by choosing values for $\bar{f}_1$ and $\hat{\omega}$.  With these we know the values of the fields at some small $\bar{r}_\text{min}$ through (\ref{inner BC}) and (\ref{inner BC 2}).  We can then integrate the solution outward from $\bar{r}_\text{min}$ using the system of equations in (\ref{static eqs}) (but with $\sigma$ and $\bar{\omega}$ replaced with $\hat{\sigma}$ and $\hat{\omega}$) and determine the value of the fields at some large $\bar{r}$.  In general, the integrated solution at large $\bar{r}$ will not equal the outer boundary conditions $\bar{f}, \bar{g}\rightarrow 0$.  We thus vary the constants $\bar{f}_1$ and $\hat {\omega}$ until it does.  Once the outer boundary conditions are satisfied, we have found a static solution.  Given a static solution, the asymptotic value of $\bar{m}$ is the ADM mass $\overline{M} = M(\mu/m_\text{Pl}^2)$ and the asymptotic value of $1/\hat{\sigma}$ is $\sigma_0$, from which $\bar{\omega} = \sigma_0 \hat{\omega}$.  

We mentioned in the Introduction that we only consider the ground state in this work.  Excited states are static solutions with nodes, which are points where $\bar{f}, \bar{g} = 0$ other than at $\bar{r}\rightarrow \infty$.  A static solution with $n$ nodes is said to be in the $n$th excited state.

\begin{figure*}
\centering
\includegraphics[width=6.5in]{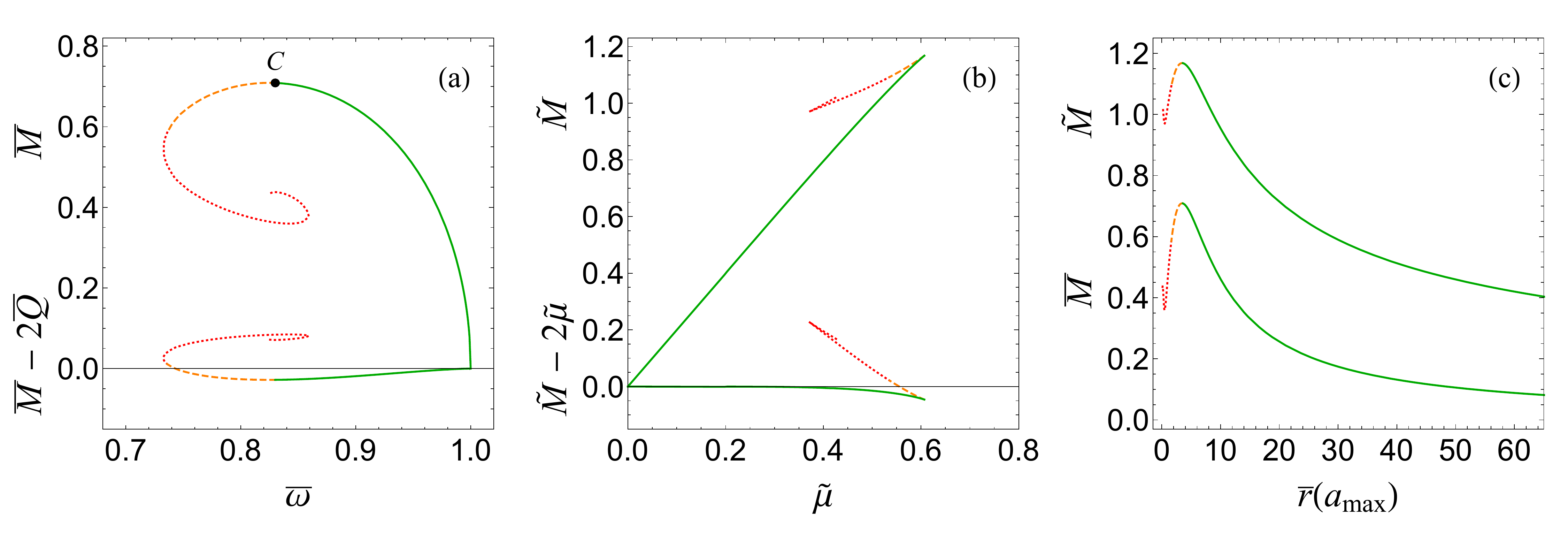}
\caption{The space of static Dirac stars is displayed in various ways.  Each curve has a one-to-one correspondence with every other curve and, in this sense, all curves are equivalent.  Every point on a curve represents a static solution.  $\overline{M}$ is the ADM mass, $\overline{M}-2\overline{Q}$ is the binding energy, and $\bar{\omega}$ is the oscillation frequency.  $\widetilde{M}$, $\widetilde{M}-2\tilde{\mu}$, and $\tilde{\mu}$ are the ADM mass, binding energy, and fermion mass when we restrict to the single particle condition $Q=1$.  $\bar{r}(a_\text{max})$ is the location where the metric function $a$ has its maximum value.  The $S$-branch is colored (solid) green and the $U$-branch is colored both (dashed) orange and (dotted) red.  The Dirac star with the largest mass is called the critical solution and is labeled $C$ in the top curve in (a) (and occurs on all curves where green turns to orange).  The value of all plotted quantities for the critical solution is given in Table \ref{table}.}
\label{fig:static space}
\end{figure*}

Figure \ref{fig:static space} displays the space of static solutions in a number of different ways.  All of the curves in Fig.\ \ref{fig:static space} are equivalent, in that every curve has a one-to-one correspondence with every other curve.  The points that make up the curves represent static solutions.  Some, though not all, of the curves have been presented elsewhere.  The top curves in \ref{fig:static space}(a) and \ref{fig:static space}(b) reproduce figures in \cite{Herdeiro:2017fhv}.  The top curve in \ref{fig:static space}(a) is a straightforward presentation of the space of static solutions, showing the ADM mass, $\overline{M}$, as a function of the frequency $\bar{\omega}$.  We see that static solutions only exist in the range $0.733 < \bar{\omega} < 1$.  Since $\bar{\omega} = \omega/\mu$, solutions only exist for $\omega < \mu$.  There is reason to believe that the spiral structure continues indefinitely, well past what we have plotted \cite{Finster:1998ws}.  The top curve in \ref{fig:static space}(b) is similar in that it plots the ADM mass, $\widetilde{M}$, but fixes $Q=1$.  We see that with $Q=1$, static solutions exist only for $0 <\tilde{\mu} < 0.607$, where $\tilde{\mu} = \mu/m_\text{Pl}$.

The bottom curves in Figs.\ \ref{fig:static space}(a) and \ref{fig:static space}(b) display the binding energy (the bottom curve in \ref{fig:static space}(b) reproduces a figure in \cite{Finster:1998ws}):\
$M$ is the total energy in the system and $2 Q \mu$ is the total rest energy, since $Q$ gives the fermion number for each type of fermion.  Their difference gives the total kinetic and gravitational energy or binding energy.  Scaling gives $\overline{M} - 2\overline{Q}$ and setting $Q=1$ gives $\widetilde{M}-2\tilde{\mu}$.  A necessary but insufficient condition for stability of a static solution is that the binding energy is negative.  The transition from negative to positive binding energy occurs at $\bar{\omega} = 0.743$ and $\tilde{\mu} = 0.555$.  

The two curves in Fig.\ \ref{fig:static space}(c) display the ADM mass ($\overline{M}$ and $\widetilde{M}$) as a function of $\bar{r}(a_\text{max})$, where $\bar{r}(a_\text{max})$ is the value of $\bar{r}$ where the metric function $a(r)$ has its maximum.  We will make heavy use of the bottom curve in \ref{fig:static space}(c) when studying stability in the next section.  The curves in \ref{fig:static space}(c) are analogous to curves used in stability studies of boson stars \cite{Seidel:1990jh, Balakrishna:1997ej} and oscillatons \cite{Alcubierre:2003sx, UrenaLopez:2012zz}.

Finster et al.\ studied the stability of these static solutions semi-analytically \cite{Finster:1998ws}, finding them stable over the portion of the curves in Fig.\ \ref{fig:static space} that we have colored (solid) green and unstable otherwise.  Though they did not mention it in \cite{Finster:1998ws}, the transition point from stable to unstable occurs where the ADM mass ($\overline{M}$ or $\widetilde{M}$) takes its maximum value, which is where green turns to (dashed) orange in all curves in Fig.\ \ref{fig:static space}.  This connects well with boson stars \cite{Gleiser:1988ih, Lee:1988av, Seidel:1990jh, Balakrishna:1997ej}, oscillatons \cite{Alcubierre:2003sx, UrenaLopez:2012zz}, and Proca stars \cite{Brito:2015pxa, Sanchis-Gual:2017bhw}, for they too transition from stable to unstable at the solution with the largest mass. 

\begin{table} 
\begin{tabular}{c|ccccccc}
$\,$&
$\bar{\omega}$ & 
$\tilde{\mu}$ & 
$\overline{M}$ & 
$\widetilde{M}$ & 
$\overline{M}-2\overline{Q}$ &
$\widetilde{M} -2\tilde{\mu}$ &
$\bar{r}(a_\text{max})$ \\
\hline
$C$ & 0.830 & 0.607 & 0.709 & 1.168 & $-0.0279$  & $-0.0459$ & 3.579 \\
BE & 0.743 & 0.555 & 0.617 & 1.111 & 0 & 0 & 1.842
\end{tabular}
\caption{The row for $C$ gives the value of all plotted quantities in Fig.\ \ref{fig:static space} for the critical solution.  The row for BE gives the value of all plotted quantities in Fig.\ \ref{fig:static space} for when the binding energy vanishes, where the binding energy is plotted in the bottom curves of Figs.\ \ref{fig:static space}(a) and (b).}
\label{table}
\end{table} 

The stability analysis in \cite{Finster:1998ws} was semi-analytical and based on first order perturbations of the static solutions.  For this reason it could not treat large perturbations nor the evolution of growing instabilities.  In the next section we dynamically evolve Dirac stars with both small and large perturbations.  We corroborate that the green portions of the curves in Fig.\ \ref{fig:static space}, which, as mentioned, extend to the largest mass solution, are stable.  We refer to the green portion as the $S$-branch.  The remainder of the curves represent intrinsically unstable static solutions and we refer to this portion of the curves as the $U$-branch.  We have colored the $U$-branch both (dashed) orange and (dotted) red for the following reason.  In the next section we will show that solutions on the orange portion of the $U$-branch, although unstable, when dynamically evolved migrate to the $S$-branch.  It may very well be that solutions on the red portion of the $U$-branch also migrate to the $S$-branch when dynamically evolved, but, as we will see in the next section, it is challenging to determine whether the solution migrates to the $S$-branch or dissipates to infinity.  The red portion of the $U$-branch, then, is that portion of the $U$-branch that we do not study dynamically.

We refer to the largest mass solution as the critical solution and the largest mass as the critical mass.  We marked the critical solution only on the top curve in Fig.\ \ref{fig:static space}(a) (the critical solution lies at the transition between the green and orange portions on all curves) and give the values of all plotted quantities for the critical solution in Table \ref{table}.  For completeness, also in Table \ref{table} is the value of all plotted quantities for when the binding energy equals zero.

\begin{figure*}
\centering
\includegraphics[width=6.5in]{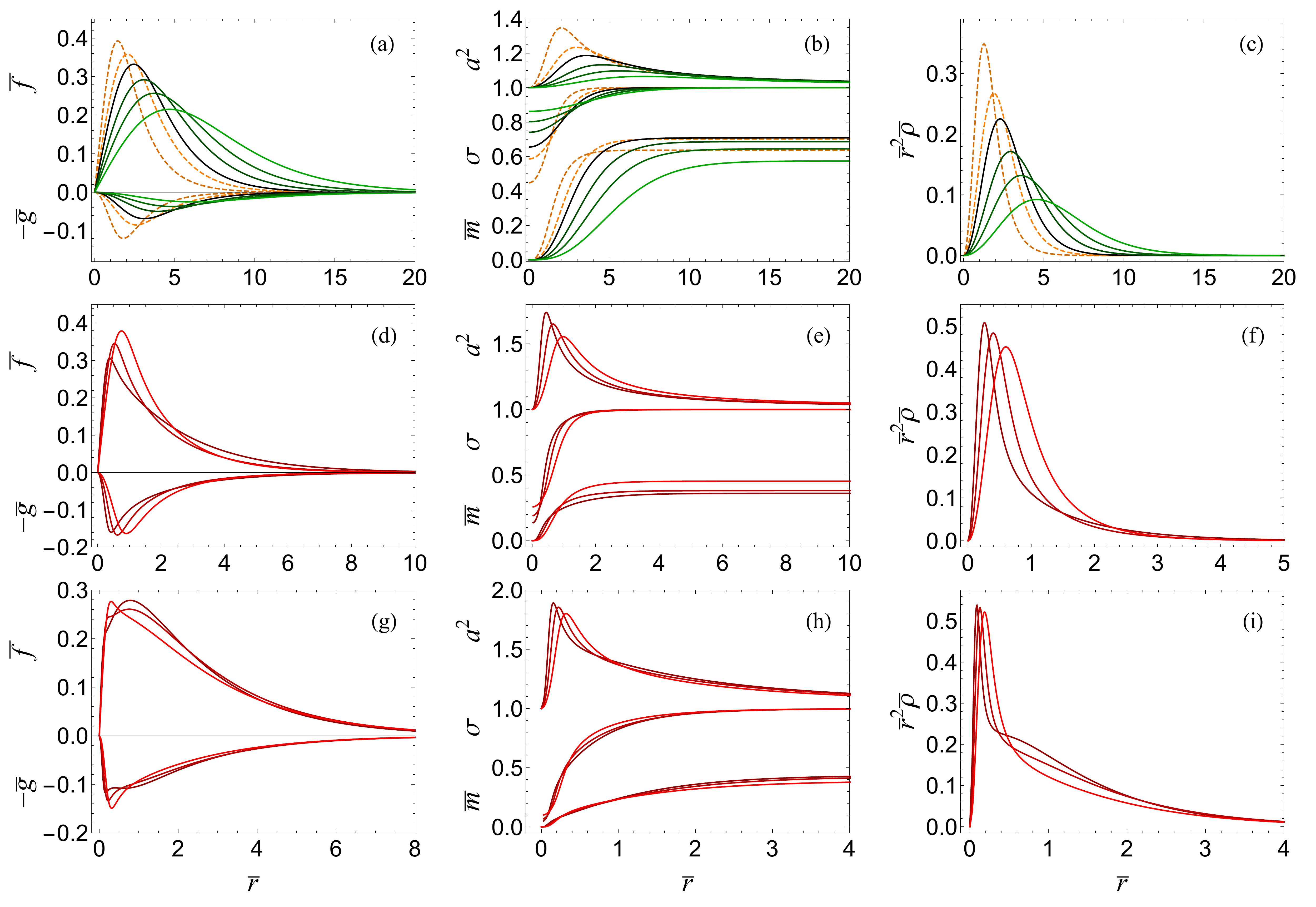}
\caption{Various static solutions are plotted.  The left column plots the fermion functions $\bar{f}$ and $\bar{g}$, the middle column plots the metric functions $a^2$, $\sigma$, and $\bar{m}$, and the right column plots the radial energy density $\bar{r}^2\rho = 4\pi r^2 \rho/m_\text{Pl}^2$.  The top row displays (solid green) $S$-branch and (dashed orange) $U$-branch solutions.  From bottom to top in $a^2$, the solutions displayed are $\bar{\omega} = 0.93$, 0.9, 0.87, 0.83, 0.8, 0.75, where $\bar{\omega} = 0.83$ is the critical solution and is colored black.  The middle row displays $U$-branch solutions from the bottom (dotted red) branch in the top curve of Fig.\ \ref{fig:static space}(a). From bottom to top in $a^2$, the solutions displayed are $\bar{\omega} = 0.75$, $0.8$, and $0.85$.  The bottom row displays $U$-branch solutions from the middle (dotted red) branch in the top curve of Fig.\ \ref{fig:static space}(a).  From bottom to top for $a^2$, the solutions displayed are $\bar{\omega} = 0.858$, $0.844$, and $0.83$.  (For consistency, all curves in the middle and bottom rows should be dotted.  We refrained from doing so to improve clarity of the curves.)  }
\label{fig:static examples}
\end{figure*}

We end this section by displaying individual solutions in Fig.\ \ref{fig:static examples}.  The top row of Fig.\ \ref{fig:static examples} displays solutions for the (solid green) $S$-branch and (dashed orange portion of the) $U$-branch and includes the critical solution.  The middle row displays $U$-branch solutions from the bottom (dotted red) branch in the top curve of Fig.\ \ref{fig:static space}(a).  The bottom row displays additional $U$-branch solutions from the middle (dotted red) branch in the top curve of Fig.\ \ref{fig:static space}(a).  The left column gives the matter functions $\bar{f}$ and $\bar{g}$, both of which are positive.  The center column shows various metric functions:\ $a^2=g_{rr}$, $\sigma$, and $m$.  The right column gives the radial energy density $\bar{r}^2 \bar{\rho} = 4\pi r^2 \rho/m_{\text{Pl}}^2$.

A couple comments are in order.  In Fig.\ \ref{fig:static examples}(c) we can see the radial energy density flattening out and its peak moving to larger $\bar{r}$ as $\bar{\omega} \rightarrow 1$.  This same flattening can be seen for $\bar{f}$, $\bar{g}$, and $a^2$ as well as the ADM mass (i.e.\ the large $\bar{r}$ limit of $\bar{m}$) decreasing in Figs.\ \ref{fig:static examples}(a) and (b).  Solutions in the limit $\bar{\omega}\rightarrow 1$ are referred to as ``dilute."

We can also see in Fig.\ \ref{fig:static examples} that as we move from the $S$-branch to the $U$-branch the peak of $a^2$ grows and the lapse, $\alpha$, which is equal to $\sigma$ at the origin, drops toward zero.  Both of these are indicators for these Dirac stars being on the verge of collapsing to form a black hole.  We shall see that this is the case in the next section.


\section{Dynamic Solutions}
\label{sec:dynamic}

In the previous section we gave a brief discussion of stability for static solutions.  In this section we make a detailed numerical investigation of stability by dynamically evolving perturbed static solutions.  We begin with a brief summary of our findings.  As mentioned in the Introduction, our investigation uses the framework of Seidel and Suen \cite{Seidel:1990jh} and our results are similar to that found in studies of boson stars \cite{Seidel:1990jh, Balakrishna:1997ej}, oscillatons \cite{Alcubierre:2003sx, UrenaLopez:2012zz}, and Proca stars \cite{Sanchis-Gual:2017bhw}.  We identify stable ($S$) and unstable ($U$) branches of static solutions.  When an $S$-branch Dirac star is given a small perturbation and then evolved, it oscillates about the original solution.  When it is evolved after being given a large perturbation that does not increase its ADM mass above the critical mass, the system migrates to the $S$-branch.  If, on the other hand, the perturbation increases the ADM mass above the critical mass, there are two possibilities:\ the system can either migrate to the $S$-branch or collapse and form a black hole. 

When a $U$-branch Dirac star is given a small perturbation, it migrates to the $S$-branch.  For large perturbations, if the perturbation increases its mass, even by as little as a few percent, it collapses to form a black hole, while if it decreases its mass, it again migrates to the $S$-branch.

In the next subsection we describe the code we use to dynamically evolve the Einstein-Dirac system.  In Sec.\ \ref{sec:perturb S} we give results for small and large perturbations of $S$-branch Dirac stars and in Sec.\ \ref{sec:perturb U} we do the same for $U$-branch Dirac stars.


\subsection{Numerics}

The matter fields to solve for are $F_1(t,r)$, $F_2(t,r)$, $G_1(t,r)$, and $G_2(t,r)$, which obey the (time-dependent) evolution equations in (\ref{matter eom}).  The metric functions to solve for are $a(t,r)$ and $\alpha(t,r)$, which obey the (time-dependent) constraint equations in (\ref{alpha a eqs}).  Since we are solving these equations numerically, we move to the dimensionless variables defined in (\ref{first scaling}).

Dynamical solutions require both inner and outer boundary conditions.  For inner boundary conditions we use $a(t,0) = 1$, which is the flat space value $a$ has when inside a spherically symmetric matter distribution.  A look at the constraint equation for $\alpha$ in (\ref{alpha a eqs}) shows that $\alpha$ can be multiplied by any constant and still be a solution.  We set $\alpha(t,0) = 1$, but after it has been solved for on a time slice we scale it so that $\alpha(t,r) = 1/a(t,r)$ at the outer boundary, since we are assuming the spacetime is asymptotically Schwarzschild.  We take the parity of $a$ and $\alpha$ to be even near the origin.  To determine the inner boundary conditions for the matter fields, we again plug in power series expansions (but this time with time-dependent coefficients) into the system of equations (\ref{matter eom}) and (\ref{alpha a eqs}).  We find that near the origin $F_{1,2} = O(r)$ and is odd and $G_{1,2} = O(r^2)$ and is even.

The outer boundary of our computational domain does not extend to $r=\infty$ and thus we need to allow fields to exit the computational domain smoothly.  We adopt the standard practice of assuming spacetime is flat at the outer boundary and thus ignore possible back-scattering from the curvature of spacetime there.  We begin by taking the large $r$ limit of the matter equations in (\ref{matter eom}) and approximating $a=\alpha = 1$.  Taking a time derivative of the resulting set of equations and then combining them we find
\begin{equation}
\begin{split}
\partial^2_{\bar{t}} \overline{F}_1 &= \partial^2_{\bar{r}} \overline{F}_1, \qquad
\partial^2_{\bar{t}} \overline{F}_2 = \partial^2_{\bar{r}} \overline{F}_2,
\\
\partial^2_{\bar{t}} \overline{G}_1 &= \partial^2_{\bar{r}} \overline{G}_1, \qquad
\partial^2_{\bar{t}} \overline{G}_2 = \partial^2_{\bar{r}} \overline{G}_2,
\end{split}
\end{equation}
which are one-dimensional (non-spherical) wave equations.  Assuming only outgoing waves exist at the outer boundary, our outer boundary conditions are \cite{Ventrella:2003fu}
\begin{equation}
\begin{split}
\dot{\overline{F}}_1 &= - \overline{F}_1', \qquad
\dot{\overline{F}}_2 = - \overline{F}_2',
\\
\dot{\overline{G}}_1 &= - \overline{G}_1', \qquad
\dot{\overline{G}}_1 = - \overline{G}_2',
\end{split}
\end{equation}
where dots and primes now denote $\bar{t}$- and $\bar{r}$-derivatives.

We composed second order accurate code to dynamically solve the massive Einstein-Dirac system.  We solve the constraint equations in (\ref{alpha a eqs}) using second order Runge-Kutta and the evolution equations in (\ref{matter eom}) using the method of lines and third order Runge-Kutta.  To improve stability we use centered sixth order finite differencing for all spatial derivatives and include fourth order Kreiss-Oliger dissipation in the evolution equations \cite{AlcubierreBook}.  Unless otherwise noted, all results presented are made with grid spacing $\Delta \bar{r} = 0.01$, an outer boundary at $\bar{r}_\text{max} = 100.005$, a time step of $\Delta \bar{t} / \Delta \bar{r} = 0.5$, and evolutions run up until $\bar{t} = 10^4$.

Determining stability dynamically requires particularly long run-times.  With an outer boundary at about $\bar{r}_\text{max} = 100$ there will inevitability be reflections.  We have tested our results by extending the outer boundary and found reflections to have a nonzero but small and mostly negligible effect on results.  In Appendix \ref{sec:code tests} we discuss various tests of our code.

\begin{figure*}
\centering
\includegraphics[width=6.5in]{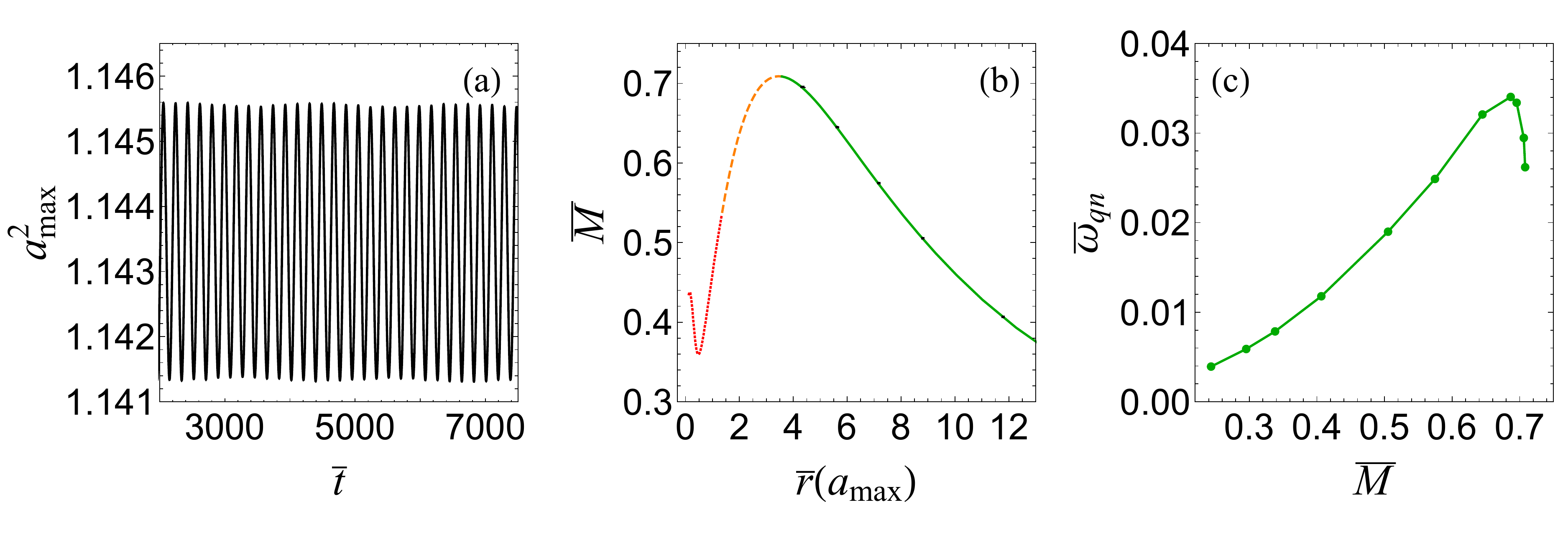}
\caption{Results for weakly perturbed $S$-branch Dirac stars. (a) shows the quasinormal oscillations of the maximum value of the metric function $a^2=g_{rr}$ for a Dirac star with $\bar{\omega} = 0.86$.  The curve in (b) is the same curve on the bottom of Fig.\ \ref{fig:static space}(c).  The black dots are time evolutions of various $S$-branch solutions.  That they do not move indicates stability.  (c) The dots give quasinormal frequencies, $\bar{\omega}_{qn}$, as a function of the ADM mass, $\overline{M}$, for evolutions of various $S$-branch solutions.  That the dots stay put is another indication of stability.  $S$-branch solutions exist for $0.830 < \bar{\omega} < 1$ and $\bar{\omega}$ increases from right to left along the curve.}
\label{fig:S branch small}
\end{figure*}


\subsection{Perturbations of $S$-branch solutions}
\label{sec:perturb S}

Figure \ref{fig:static space} displays the space of static solutions.  We colored (solid) green the portion of the curves in Fig.\ \ref{fig:static space} which represent stable solutions and we refer to the green portion as the $S$-branch.  Figure \ref{fig:static examples} displays a few $S$-branch solutions.  In this subsection we use perturbed $S$-branch solutions as initial data and then dynamically evolve them forward in time.  We begin with weak perturbations before moving to strong perturbations.  As is well known, discretization error inherent in numerical solutions acts as a weak perturbation \cite{Seidel:1990jh, Balakrishna:1997ej, Alcubierre:2003sx, UrenaLopez:2012zz, Sanchis-Gual:2017bhw}.

In Fig.\ \ref{fig:S branch small}(a) we display a time evolution of $a^2_\text{max}$, the maximum value of the metric function $a^2 = g_{rr}$, for the weakly perturbed $S$-branch solution with $\bar{\omega} = 0.86$.   Figure \ref{fig:S branch small}(a) clearly shows oscillations, but it also demonstrates stability in that it is absent of decay or any major change in the configuration of the solution.  The oscillations in Fig.\ \ref{fig:S branch small}(a) are called quasinormal modes and we use $\bar{\omega}_{qn}$ to label the quasinormal frequency of these oscillations.  Figure \ref{fig:S branch small}(a) is typical in that evolutions of other weakly perturbed $S$-branch solutions give similar results.

A different way to understand stability is shown in Fig.\ \ref{fig:S branch small}(b).  The curve is the same as in Fig.\ \ref{fig:static space}(c) and displays the space of static solutions.  The small black dots on the (solid green) $S$-branch are time evolutions of weakly perturbed $S$-branch solutions.  That the black dots stay in place and do not move away from the curve indicates that they are stable.  (This may not be clear now, but will be be made more clear when we consider large perturbations in a moment.)  Like Fig.\ \ref{fig:S branch small}(a), the black dots show that $S$-branch solutions are stable with respect to small perturbations.

Our final method of displaying results is shown in Fig.\ \ref{fig:S branch small}(c).  We computed the quasinormal frequencies, $\bar{\omega}_{qn}$, of weakly perturbed $S$-branch solutions by Fourier transforming the data in Fig.\ \ref{fig:S branch small}(a) and analogous data from the evolution of other $S$-branch solutions (the Fourier transform of Fig.\ \ref{fig:S branch small}(a) is given in Appendix \ref{sec:code tests}).  The ADM mass, $\overline{M}$, and the quasinormal frequencies are roughly constant over the evolution and so each $S$-branch solution has a single point in Fig.\ \ref{fig:S branch small}(c).  We see that the curve reaches a maximum before turning and heading to zero, just as it does with boson stars and oscillatons, which is an indication that the system is heading toward instability \cite{Seidel:1990jh}.  We will find below that this plot is particularly useful in understanding both strongly perturbed $S$-branch solutions and $U$-branch solutions, just as it is for boson stars and oscillatons.  The conclusion we draw from the results in Fig.\ \ref{fig:S branch small} is that $S$-branch solutions are stable with respect to weak perturbations, corroborating what was found semi-analytically in \cite{Finster:1998ws}.

We turn now to strongly perturbed $S$-branch Dirac stars.  There is great variety in the perturbations we could use, which includes their shape, their location, and which of the four fermion fields ($F_1$, $F_2$, $G_1$, and $G_2$) they perturb.  For simplicity we show results only for Gaussian perturbations applied to $F_1$:
\begin{equation} \label{F1 perturbation}
\delta \overline{F}_1(0,\bar{r}) = C (\bar{r}/\bar{r}_0) e^{-(\bar{r}-\bar{r}_0)^2},
\end{equation}
where $C$ and $\bar{r}_0$ are constants.  This is similar to what was used in \cite{Seidel:1990jh, Alcubierre:2003sx}, and like those works we will describe the perturbation in terms of its effect on the ADM mass of the system.  We have experimented with other perturbations and found they give similar results.

Our results are very similar to what is found for boson stars and oscillatons.  If the perturbation does not raise the ADM mass above the critical mass, $\overline{M}_C$ (i.e.\ the largest allowed mass for static solutions), then the system migrates to the $S$-branch.  On the other hand, if the perturbation does raise the mass above the critical mass, then there are two options:\  The system migrates to the $S$-branch or it collapses and forms a black hole.  For dilute Dirac stars, the ADM mass can be raised significantly higher than the critical mass without ending in collapse, since dilute stars are particularly efficient at ejecting mass via gravitational cooling \cite{Seidel:1993zk}.  For example, for the $\bar{\omega} = 0.98$ static solution, we were able to raise the ADM mass to 50\% above $\overline{M}_C$ and still have the system migrate back to the $S$-branch.  On the other hand, for static solutions with ADM masses close to the critical mass, raising the ADM by as little as 1\% above $\overline{M}_C$ quickly leads to collapse.

\begin{figure}
\centering
\includegraphics[width=2.25in]{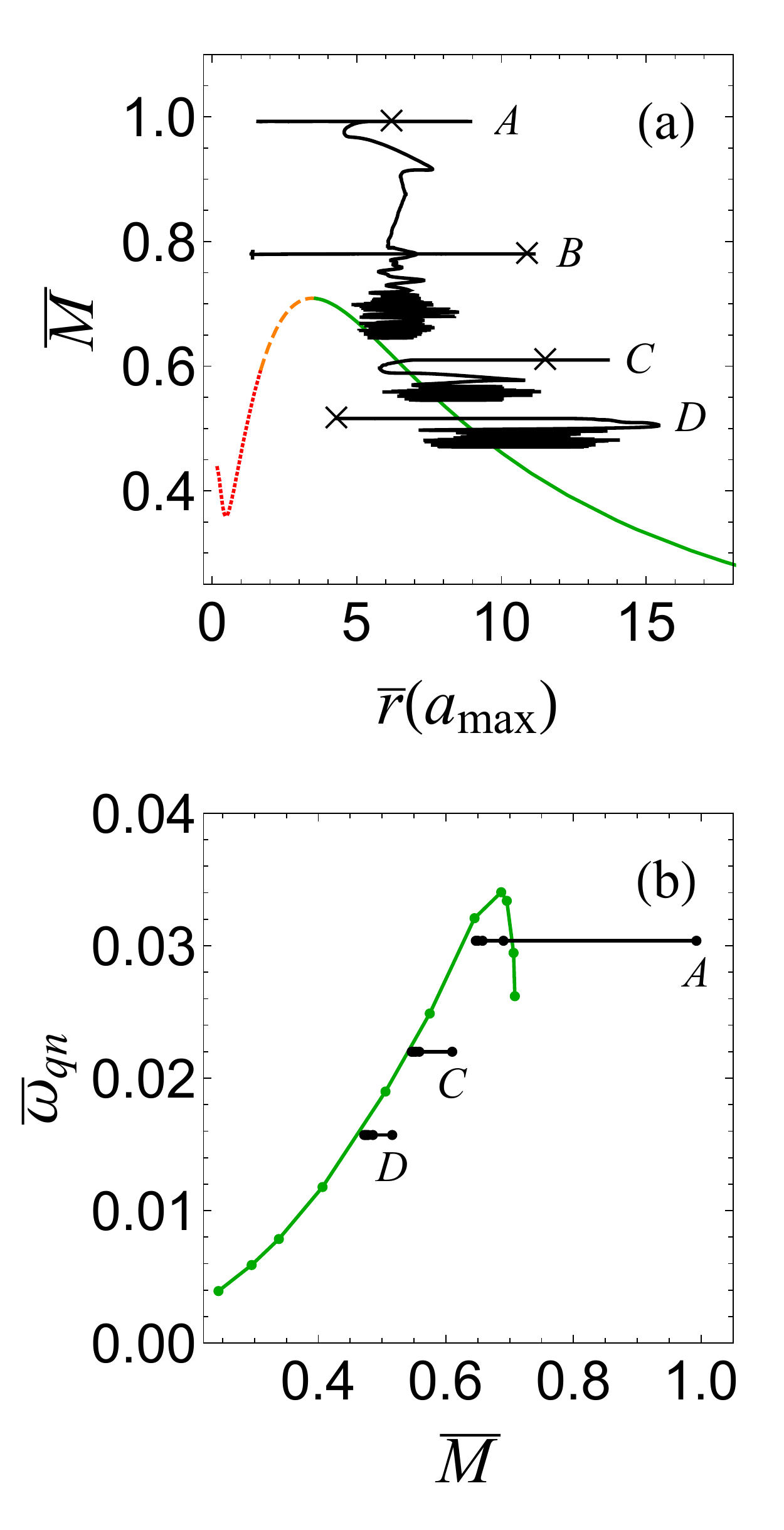}
\caption{Results for strongly perturbed $S$-branch Dirac stars.  (a)  The crosses mark the beginning of the time evolution.  Solutions $A$, $C$, and $D$ migrate to the $S$-branch while solution $B$ collapses and forms a black hole.  This may be compared with Fig.\ \ref{fig:S branch small}(b).  (b) The quasinormal frequencies are found to be fairly constant during the evolution and give an indication as to the $S$-branch solution the system is migrating to.  The black lines each have five dots, marking their ADM masses at times $\bar{t} = 0$, 2000, 4000, 6000, and 8000.  The details of the initial static solutions and the perturbations are found in the main text.}
\label{fig:S branch large}
\end{figure}

Some examples of these possibilities are shown in Fig.\ \ref{fig:S branch large}.  Figure \ref{fig:S branch large}(a) displays the evolution of four strongly perturbed $S$-branch solutions, labeled $A$, $B$, $C$, and $D$.  The crosses mark where the evolution begins.  Each point on a black curve represents a different moment in time and so the curves display how the solution is dynamically changing.  This may be compared to Fig.\ \ref{fig:S branch small}(b) (which hopefully now is more clear), where the black curves do not move because the starting configurations are stable.

Solution $A$ corresponds to a dilute Dirac star ($\bar{\omega} = 0.98$) with a perturbation that raises its mass to 40.0\% above $\overline{M}_C$ (which corresponds to raising its ADM mass by 94\%).  Solution $A$ illustrates how dilute static solutions can have their mass raised significantly higher than the critical mass and still migrate to the $S$-branch and not collapse.  Curve $C$ is the $\bar{\omega} = 0.97$ solution with its mass increased 50\% and curve $D$ is the $\bar{\omega} = 0.9$ solution with its massed decreased 20.0\%.  All three of these strongly perturbed solutions can be seen to migrate to the $S$-branch.

Curve $B$ is different in that it collapses to a black hole.  This is the $\bar{\omega} = 0.93$ solution with its mass increased to 10.0\% above $M_C$ (which corresponds to increasing its mass 35.7\%).  The $B$ curve moves to the left edge, where $\bar{r}(a_\text{max})$ stays still at the horizon radius after the black hole forms.  To determine if a black hole forms we use the standard method of looking for a sharp spike in the metric function $a$ and the collapse of the lapse, $\alpha$, in which $\alpha$ drops to zero at the origin.  Curve $B$ illustrates how static solutions with masses near the critical mass, and hence that are far from being dilute, do not require their mass to be raised much above the critical mass for them to collapse.

Like boson stars and oscillatons, we find that the quasinormal oscillation frequency stays roughly constant even as the evolving solution loses mass.  In Fig.\ \ref{fig:S branch large}(b), which again shows evolutions $A$, $C$, and $D$, the specific dots on the black lines differ in time by $\Delta \bar{t} = 2000$.  We see that the dots are very close together on the left edges, indicating that the amount of mass being lost is decreasing considerably, as is expected since they are nearing the (green) stable solution they are migrating to.  Figure \ref{fig:S branch large}(b) is further evidence that the strongly perturbed solutions do in fact migrate back to the $S$-branch.

\begin{figure*}
\centering
\includegraphics[width=6.5in]{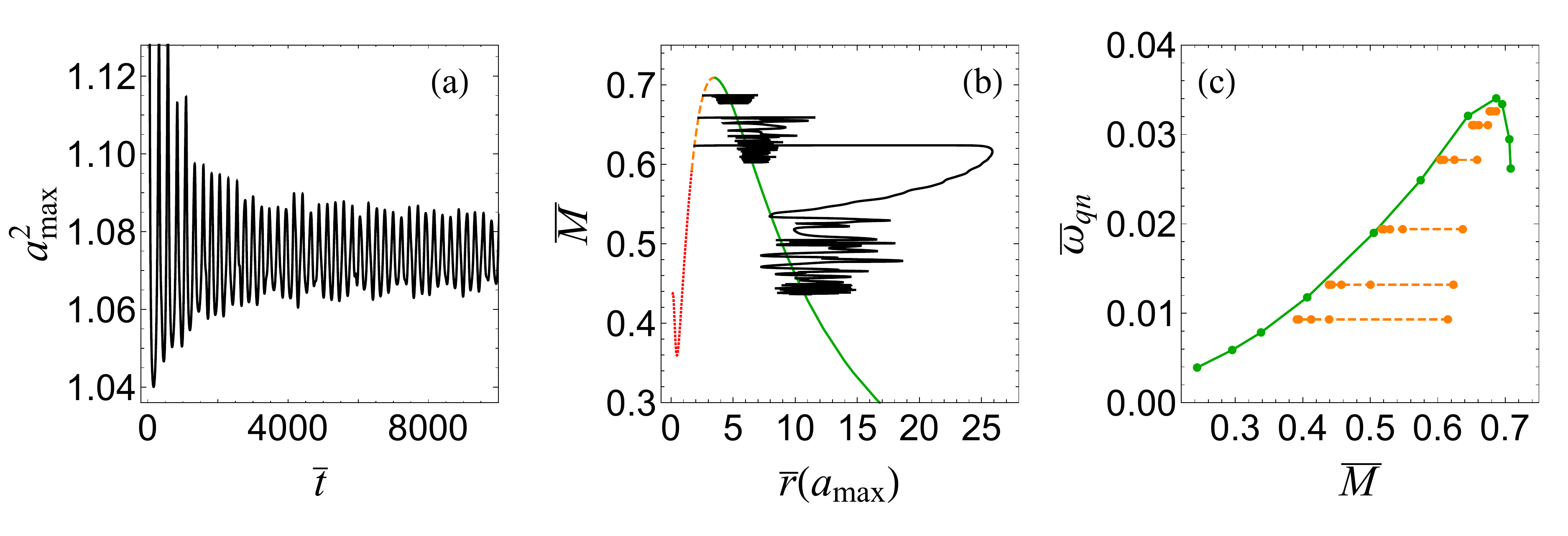}
\caption{Results for weakly perturbed $U$-branch Dirac stars.  (a) The quasinormal oscillations of the maximum value of the metric function $a^2 = g_{rr}$ for a Dirac star with $\bar{\omega} = 0.76$.  The curve in (b) is the same curve on the bottom of Fig.\ \ref{fig:static space}(c).  The black lines are time evolutions of various $U$-branch solutions.  We see that they migrate to the $S$-branch.  (c) The quasinormal frequencies as a function of their ADM mass for various $U$-branch solutions.  The quasinormal frequencies are found to be fairly constant during the evolution and give an indication as to the $S$-branch solution the system is migrating to.  The (dashed) orange lines each have five dots, marking their ADM masses at times $\bar{t} = 0$, 2000, 4000, 6000, and 8000.}
\label{fig:U branch small}
\end{figure*}

\subsection{Perturbations of $U$-branch solutions}
\label{sec:perturb U}

The $U$-branch is the (dashed) orange and (dotted) red portions of the curves in Fig.\ \ref{fig:static space}.  It is defined as the set of static solutions that are intrinsically unstable.  Figure \ref{fig:static examples} displays a number of $U$-branch solutions.  Figure \ref{fig:U branch small} displays results for weakly perturbed $U$-branch solutions that are analogous to Fig.\ \ref{fig:S branch small} for the $S$-branch.  Figure \ref{fig:U branch small}(a) displays a time evolution of $a^2_\text{max}$ for a $U$-branch solution with $\bar{\omega} = 0.76$.  The oscillations clearly change amplitude at early times before settling into seemingly stable oscillations.  Figure \ref{fig:U branch small}(b) shows the time evolution of three weakly perturbed $U$-branch solutions.  The three black curves each begin on the (dashed orange) $U$-branch.  Each point on the black curves represents a different moment in time during the evolution.  This plot indicates that the $U$-branch solutions, when weakly perturbed, migrate to the $S$-branch.  In Fig.\ \ref{fig:U branch small}(c) we show quasinormal frequencies versus ADM masses for various evolutions.  The (solid) green curve is from the $S$-branch.  The (dashed) orange lines display the weakly perturbed $U$-branch solutions of this subsection.  As before, the quasinormal frequencies of weakly perturbed $U$-branch solutions are roughly constant over the entire evolution.  As the solutions evolve, they shed mass.  The specific dots on the orange lines in Fig.\ \ref{fig:U branch small}(c) differ in time by $\Delta \bar{t} = 2000$.  We see that the dots are very close together on the left edges, indicating that the amount of mass being lost is decreasing considerably, as is expected since they are nearing the (green) stable solution they are migrating to.

\begin{figure}
\centering
\includegraphics[width=2.5in]{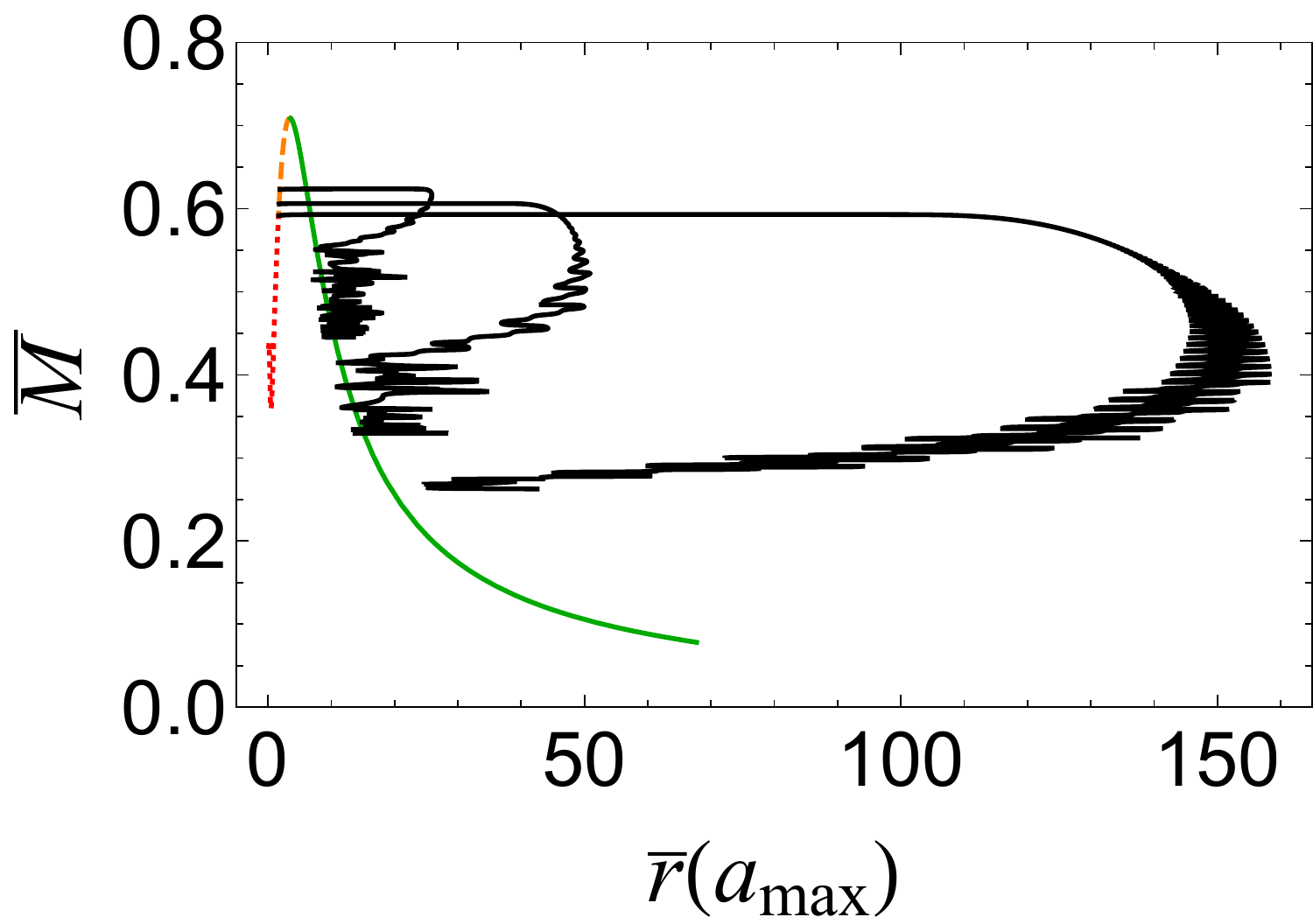}
\caption{Evolving weakly perturbed $U$-branch solutions becomes increasingly difficult as we move further along the $U$-branch.  The (dotted) red portion of the $U$-branch is that portion that we do not dynamically study and thus we cannot say whether such $U$-branch Dirac stars migrate to the $S$-branch or dissipate to infinity.}
\label{fig:U branch red}
\end{figure}

Figure \ref{fig:U branch small} shows results only from the (dashed) orange portion of the $U$-branch.  The reason is that outside this portion, it becomes numerically challenging to determine whether a weakly perturbed $U$-branch solution migrates to the $S$-branch or dissipates to infinity.  What we mean by this can be seen in Fig.\ \ref{fig:U branch red}, where we display results for static solutions that are farther along the $U$-branch (we extended the outer boundary of our computational grid to make this figure).  We can see that the peak of the metric function $a$ travels an increasingly farther distance before turning around and migrating toward the $S$-branch.  Moreover, the migration is toward increasingly more dilute regions of the $S$-branch.  Dirac stars farther along the $U$-branch are then too challenging for us to determine whether they migrate to the $S$-branch or dissipate to infinity.  As mentioned previously, the portion of the $U$-branch that we have colored (dotted) red is the portion that we have not studied dynamically.

\begin{figure}
\centering
\includegraphics[width=2.25in]{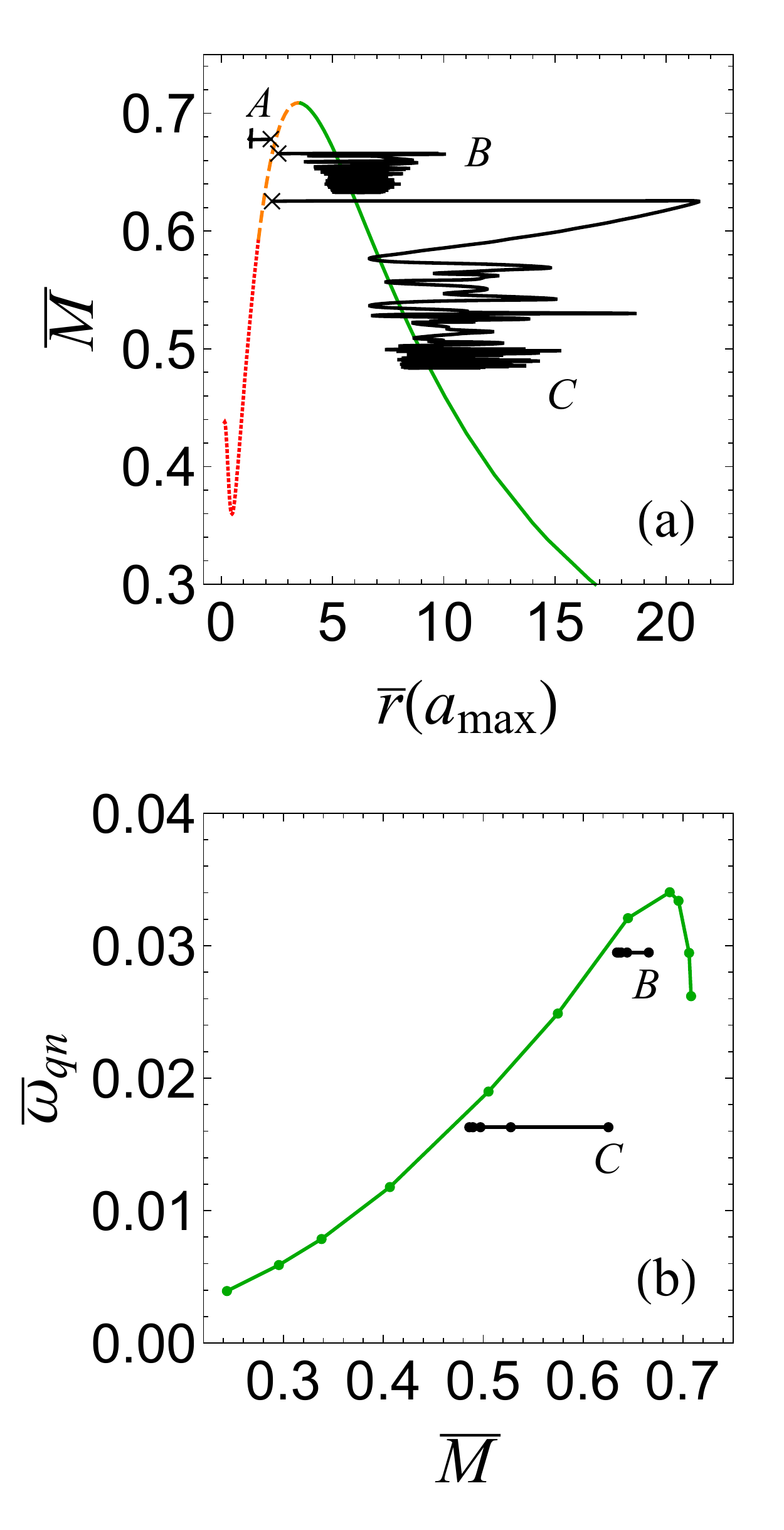}
\caption{Results for strongly perturbed $U$-branch Dirac stars.  The crosses in (a) mark the beginning of the time evolution.  Solutions $B$ and $C$ migrate to the $S$-branch while solution $A$ collapses and forms a black hole.  (b) The quasinormal frequencies are found to be fairly constant during the evolution and give an indication as to the $S$-branch solution the system is migrating to.  Further details are the same as found in Fig.\ \ref{fig:S branch large}.}
\label{fig:U branch large}
\end{figure}

We turn now to strongly perturbed $U$-branch solutions.  We again show results for the perturbation in (\ref{F1 perturbation}) and describe the perturbation in terms of its effect on the ADM mass of the system.  Similar to boson stars and oscillatons, if the perturbation decreases the ADM mass, the system migrates to the $S$-branch, while if the perturbation increases the mass even as little as a few percent, independent of whether the mass is above or below the critical mass, the system collapses and forms a black hole.  Some examples of these possibilities are shown in Fig.\ \ref{fig:U branch large}.  Figure \ref{fig:U branch large}(a) displays the evolution of three strongly perturbed $U$-branch solutions, labeled $A$, $B$, and $C$.  Solution $B$ corresponds to a $U$-branch Dirac star with $\bar{\omega} = 0.78$ and a perturbation that decreases its mass by 3.0\% and solution $C$ corresponds to $\bar{\omega} = 0.76$ with a perturbation that decreases its mass by 5.0\%.  Both strongly perturbed solutions can be seen to migrate to the $S$-branch.  Curve $A$ is different in that it collapses to form a black hole.  This is the $\bar{\omega} = 0.74$ solution with a perturbation that increases its mass by 3.0\%.  As before, we have found that the migrating solutions have quasinormal frequencies that stay roughly constant during the evolution.  We show these in Fig.\ \ref{fig:U branch large}(b) as further evidence that strongly perturbed $U$-branch solutions that do not collapse will migrate to the $S$-branch.


\section{Conclusion}
\label{sec:conclusion}

Dirac stars are self-gravitating systems of spin-1/2 fermions as described by the Dirac equation.  Static solutions were found by Finster et al.\ in \cite{Finster:1998ws}, who made a semi-analytical stability analysis, identifying stable ($S$) and unstable ($U$) branches for their solutions.  We made a dynamical study of stability of spherically symmetric Dirac stars using the framework of Seidel and Suen \cite{Seidel:1990jh}.  This framework has been used to study the stability of boson stars \cite{Seidel:1990jh, Balakrishna:1997ej} and oscillatons \cite{Alcubierre:2003sx, UrenaLopez:2012zz}, with a similar framework used for Proca stars \cite{Sanchis-Gual:2017bhw}.

We corroborated that weakly perturbed $S$-branch Dirac stars are stable and found that strongly perturbed $S$-branch Dirac stars will migrate to the $S$-branch if their mass is not increased above the critical mass, which is the largest mass found among the static solutions.  If, on the other hand, a strongly perturbed $S$-branch Dirac star does have its mass increased beyond the critical mass, then it can either migrate to the $S$-branch or collapse and form a black hole.  For $U$-branch Dirac stars, we found that weakly perturbed solutions and strongly perturbed solutions that have their mass decreased migrate to the $S$-branch, while increasing the mass even by as a little as a few percent leads to collapse.

We end by commenting on limitations of our work.  Our study is entirely classical, just as with \cite{Ventrella:2003fu, Dzhunushaliev:2018jhj, Dzhunushaliev:2019kiy}.  It remains to be determined how important quantum effects are to Dirac stars \cite{Herdeiro:2017fhv, Blazquez-Salcedo:2019qrz}.  We only considered Dirac stars in the ground state.  Intriguingly, the excited states are known to have a stable branch \cite{Finster:1998ws}.  A dynamical stability analysis of the excited states similar to what has been done for boson stars \cite{Balakrishna:1997ej} may have interesting results.  Such a study will be presented elsewhere.  We ignored self-interactions, which were shown in \cite{Dzhunushaliev:2018jhj, Dzhunushaliev:2019kiy} to have important astrophysical consequences.  This too may benefit from a stability analysis similar to what was done for boson stars \cite{Balakrishna:1997ej}.  Finally, we studied spherically symmetric solutions.  Solutions for spinning Dirac stars were recently found \cite{Herdeiro:2019mbz}.  A stability analysis for such solutions, which has recently been initiated for spinning boson and Proca stars \cite{Sanchis-Gual:2019ljs}, would be interesting.


\begin{acknowledgments}
E.\ D.\ thanks Janna L.\ Murgia-Hoppin and John W.\ Hoppin, Ph.D.\ for financial support.  N.\ N.\ P.\ thanks the Holy Cross Alumni/Parents Summer Research Scholarship Fund for financial support.
\end{acknowledgments}


\appendix

\section{Relation of spinor ansatz to literature}
\label{app:ansatz}

The spherically symmetric Dirac spinor ansatz we presented in (\ref{psi pm}) does not have the same form as found elsewhere in the literature \cite{Finster:1998ws, Ventrella:2003fu, Herdeiro:2017fhv} but, as we show here, it is equivalent.  The first group to find static solutions in the Einstein-Dirac system was Finster, Smoller, and Yau (FSY) \cite{Finster:1998ws}.  FSY described their static fermions in terms of the two real functions $\alpha_\text{FSY}(r)$ and $\beta_\text{FSY}(r)$.  Our static fermion functions $f(r)$ and $g(r)$ are related to theirs through
\begin{equation}
\alpha_\text{FSY} = \left( \frac{\alpha}{4\pi a} \right)^{1/2} f, \qquad
\beta_\text{FSY} = \left( \frac{\alpha}{4\pi a} \right)^{1/2} g.
\end{equation} 
With this relation (and the simple relation between our and their metric functions) our static equations are identical to theirs.

A different Dirac spinor ansatz for static solutions was used by Herdeiro, Pombo, and Radu (HPR) \cite{ Herdeiro:2017fhv}.  HPR described their static fermions in terms of the functions $f_\text{HPR}(r)$ and $g_\text{HPR}(r)$.  Using
\begin{equation}
f_\text{HPR} = \frac{g}{4r \sqrt{\pi a}}, \qquad
g_\text{HPR} = \frac{f}{4r \sqrt{\pi a}}
\end{equation}
our static equations are identical to theirs.

Ventrella and Choptuik (VC) \cite{Ventrella:2003fu} were the first authors to dynamically solve the Einstein-Dirac system, but only for massless fermions.  VC describe their dynamic fermions in terms of the functions $F_1^\text{VC}(t,r)$, $F_2^\text{VC}(t,r)$, $G_1^\text{VC}(t,r)$, and $G_2^\text{VC}(t,r)$.  Setting $\mu = 0$, so that our equations describe massless fermions, our dynamic fermion functions $F_1(t,r)$, $F_2(t,r)$, $G_1(t,r)$, and $G_2(t,r)$ are related to theirs through
\begin{equation}
\begin{split}
F_1^\text{VC} &= \frac{F_1 + G_1}{\sqrt{2}}, \qquad
F_2^\text{VC} = \frac{F_2 + G_2}{\sqrt{2}},
\\
G_1^\text{VC} &= \frac{G_2 - F_2}{\sqrt{2}}, \qquad
G_2^\text{VC} = \frac{F_1 - G_1}{\sqrt{2}}.
\end{split}
\end{equation}
With this relation our dynamic equations are identical to theirs.


\section{Code tests}
\label{sec:code tests}

\begin{figure*}
\centering
\includegraphics[width=6.5in]{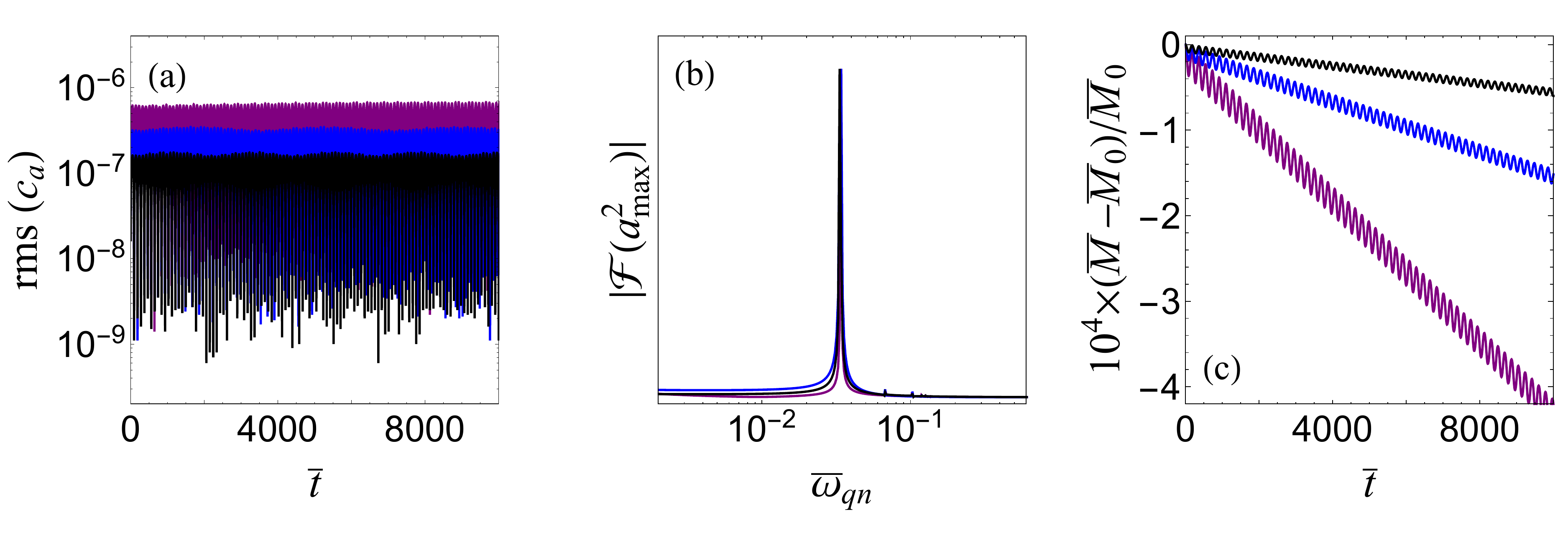}
\caption{Results for various tests we performed on our code.  In all three plots the grid spacing used to compute the results is $\Delta \bar{r} = 0.01$ (black), $0.01\!\times\!\sqrt{2}$ (blue), and $0.02$ (purple).  (a) The constraint in (\ref{c_a}) as a function of time.  That the results drop by a factor of 2 when the grid spacing drops by a factor of $\sqrt{2}$ indicates second order convergence.  (b) The grid spacing does no have an effect on the Fourier transform and our determination of the quasinormal frequency.  (c) The percent mass dissipation, where $\overline{M}$ is the total integrated mass and $\overline{M}_0$ is the initial mass.  The dissipation is negligibly small and numerical in nature.}
\label{fig:convergence}
\end{figure*}

In this appendix we present various tests of our code.  The bottom equation in (\ref{alpha a eqs}) is a constraint equation for the metric function $a(t,r)$.  The Einstein field equations also offer an evolution equation for $a(t,r)$ \cite{AlcubierreBook, BaumgarteBook},
\begin{equation} \label{a evo}
\dot{a} = -4\pi G r \alpha a S_r,
\end{equation}
where the momentum density $S_r(t,r)$ follows from the energy-momentum tensor and is given by
\begin{equation}
S_r =
\frac{1}{2\pi r^2 a} (F_1 F_2' -  F_1'F_2 + G_1 G_2' - G_1'G_2 ).
\end{equation}
Since our dynamical solutions do not make use of the above equation, it is available for code testing.  We define the constraint
\begin{equation} \label{c_a}
c_a(t,r) \equiv a_\text{code}(t,r) - a_\text{evo}(t,r),
\end{equation}
where $a_\text{code}$ is the value of $a$ found by our code and $a_\text{evo}$ is the value of $a$ found using the evolution equation (\ref{a evo}).  In Fig.\ \ref{fig:convergence}(a) we have plotted the root-mean-square (rms) of $c_a$ across the computational grid for the evolution of the $\bar{\omega} = 0.86$ $S$-branch solution for three different grid spacings:\ $\Delta \bar{r} = 0.01$, $0.01\!\times\!\sqrt{2}$, and $0.02$ (for $\Delta \bar{r} = 0.01$ this is the same evolution as shown in Fig.\ \ref{fig:S branch small}(a)).  That the results in Fig.\ \ref{fig:convergence}(a) are small indicates that the constraint $c_a = 0$ is obeyed and that the results drop by a factor of 2 when the grid spacing drops by a factor of $\sqrt{2}$ indicates second order convergence.

Figure\ \ref{fig:convergence}(b) is an example of the Fourier transform that we use to determine the quasinormal frequency $\bar{\omega}_{qn}$.  Specifically, it shows the Fourier transform of the same evolution with the same three grid spacings shown in Fig.\ \ref{fig:convergence}(a).  The vertical scale is arbitrary and the curves have been normalized so that the heights of their spikes are equal.  Figure\ \ref{fig:convergence}(b) indicates that grid spacing does not affect our determination of quasinormal frequencies.  We have also confirmed that the location of the outer boundary, and thus possible reflections at the outer boundary, does not affect our determination of quasinormal frequencies.

Finally, Fig.\ \ref{fig:convergence}(c) displays the percent change of the total integrated mass, or ADM mass, $\overline{M}$, again for the evolution and grid spacings shown in Fig.\ \ref{fig:convergence}(a).  Figure \ref{fig:convergence}(c) indicates that our code has a small amount of mass dissipation, but that the amount of dissipation is negligible and further that the dissipation is numerical in origin and not representative of a physical instability of the static solution.

We have found results similar to Fig.\ \ref{fig:convergence} for the evolutions of other static solutions, including unstable and strongly perturbed static solutions.




%

\end{document}